\documentclass[12pt, letter]{article}
\usepackage{standalone}
\usepackage{amsmath}
\usepackage{amssymb}
\usepackage{graphicx}
\usepackage{bm}
\usepackage{geometry}

\usepackage{fancyhdr}
\usepackage{caption}
\usepackage{subcaption}
\usepackage{paralist}
\usepackage{xcolor}
\usepackage[inline]{enumitem}
\usepackage{hyperref}
\usepackage{enumitem}

\usepackage{typearea}
\usepackage{adjustbox}

\usepackage{tikzit}

\tikzstyle{simple}=[fill=white, draw=black, shape=circle]

\tikzstyle{arrow}=[->, thick]

\input{example.tikzdefs}

\usepackage{tikz-network}

\usepackage{booktabs}
\usepackage{longtable}
\usepackage{array}
\usepackage{multirow}
\usepackage{wrapfig}
\usepackage{float}
\usepackage{colortbl}
\usepackage{pdflscape}
\usepackage{tabu}
\usepackage{threeparttable}
\usepackage{threeparttablex}
\usepackage[normalem]{ulem}
\usepackage{makecell}

\usepackage{rotating}

\usepackage{amsthm}
\usepackage{comment}
\usepackage{xcolor}

\theoremstyle{plain}

\newtheorem{proposition}{Proposition}

\theoremstyle{definition}

\theoremstyle{remark}

\newtheorem{example}{Example}

\usepackage{algorithm}
\usepackage{algorithmic}

\usepackage[round]{natbib}   
\newcommand{\tf}[1]{\textbf{#1}}

\newcommand{\CI}{\operatorname{CI}}

\makeatletter
\def\singlespace{\def\baselinestretch{1}\@normalsize}

\renewcommand{\baselinestretch}{1.2}
\newcommand{\lb}[1]{\textcolor{blue}{#1}}
\newcommand{\CC}[1]{\textcolor{purple}{#1}}

\renewcommand{\baselinestretch}{1.0}

\begin{document}

\title{\textbf{Causal Structural Learning on MPHIA Individual Dataset}}

\author{Le Bao$^1$, Changcheng Li$^2$, Runze Li$^1$ and Songshan Yang$^3$\\
\ 
\\
$^1$Department of Statistics, The Pennsylvania State University\\ University Park, PA 16802, USA\\
$^2$School of Mathematical Sciences, Dalian University of Technology\\ Dalian, P.R. China\\
$^3$Institute of Statistics and Big Data, Renmin University of China, \\
Beijing, P.R. China}
\date{}

\maketitle

\begin{abstract}
    The Population-based HIV Impact Assessment (PHIA) is an ongoing project that conducts nationally representative HIV-focused surveys for measuring national and regional progress toward UNAIDS’ 90-90-90 targets, the primary strategy to end the HIV epidemic. We believe the PHIA survey offers a unique opportunity to better understand the key factors that drive the HIV epidemics in the most affected countries in sub-Saharan Africa. In this article, we propose a novel causal structural learning algorithm to discover important covariates and potential causal pathways for 90-90-90 targets.
    Existing constrained-based causal structural learning algorithms are quite aggressive in edge removal. The proposed algorithm  preserves more information about important features and potential causal pathways. It is applied to the Malawi PHIA (MPHIA) data set and leads to interesting results. 
    For example, it discovers age and condom usage to be important for female HIV awareness; the number of sexual partners to be important for male HIV awareness; and knowing the travel time to HIV care facilities leads to a higher chance of being treated for both females and males.
    We further compare and validate the proposed algorithm using BIC and using Monte Carlo simulations, and show that the proposed algorithm achieves improvement in true positive rates in important feature discovery over existing algorithms.
\end{abstract}

\vspace{0.2cm} \noindent{\bf Keywords}: Causal structural learning; HIV; 90-90-90 targets; PHIA. 

\newpage

\baselineskip=22pt
\renewcommand{\baselinestretch}{1.5}

\section{Introduction}

In 2014, the United Nations Joint Programme on HIV and AIDS (UNAIDS) set the 90-90-90 targets as the primary strategy to end the HIV/AIDS epidemic \citep{joint201490}, which includes identifying 90\% of people living with HIV through expanded testing, placing 90\% of positively identified individuals on antiretroviral therapy, and ensuring that 90\% of those on therapy can achieve undetectable viral loads by 2020.
Considerable progress has been made towards Tri90 \citep{Gaolathe2016, gisslen2017sweden, joint201690, joint2017ending, marukutira2018tale}.
Yet, as of 2019, there was a significant gap; instead of the targets of 90-90-90 (Tri90), it was 81-67-59 globally \citep{joint2020good}; and now the COVID-19 crisis has the potential to undermine existing efforts towards the HIV/AIDS epidemic \citep{joint2020seizing}.



To end HIV/AIDS epidemic, we shall learn from the past efforts towards Tri90 and identify important features that could guide more targeted and effective health policies. Novel datasets and sophisticated modeling tools are needed to enhance our understanding of Tri90 achievements. The Population-based HIV Impact Assessment (PHIA) survey is a nationally representative HIV-focused survey that started data collection in 2015. It is designed to measure the reach and impact of HIV programs. We believe it offers a unique opportunity to better understand the key factors that drive the HIV epidemics in the most affected countries in sub-Saharan Africa. In this article, we propose a novel causal structural learning algorithm to discover important covariates and potential causal pathways for the Tri90. Causal structural learning aims to build a directed acyclic graph (DAG) that shows direct causal relations among variables of interest in a given domain. The resulting DAG helps us to understand the mechanism behind data.


Many classical structural learning algorithms are constrained-based, such as the PC algorithm \citep{spirtes1991algorithm}, the PC-stable algorithm \citep{colombo2014order}, and the MMPC algorithm \citep{tsamardinos2003mmpc}. 
The constrained-based algorithms learn graphical structures by d-separation set searching. 
\textcolor{black}{
D-separation is an important graphical concept for causal structural learning \citep{geiger1990logic, geiger1990d}, where the ``d" stands for dependence or directed. 
Roughly speaking, in a directed acyclic graph (DAG), two vertices $X$ and $Y$ are d-separated by a set of vertices  $\tf{Z}=\{Z_1,\cdots,Z_d\}$ if any only if all the paths/information between $X$ and $Y$ are blocked by vertices in $\tf{Z}$ (see Section~\ref{section:Preliminaries} for a rigorous definition of d-separation).
The d-separation relationship in a DAG can be related to the conditional independence through the Markov condition and the faithfulness assumption \citep{spirtes2010introduction} (see definitions in Section~\ref{section:Preliminaries}).
That is to say, vertices $X$ and $Y$ are d-separated by a vertex set $\tf{Z}=\{Z_1,\cdots,Z_d\}$ is equivalent to
the conditional independence of the corresponding variables $X$ and $Y$ given the corresponding set of variables $\tf{Z}=\{Z_1,\cdots,Z_d\}$ under the Markov condition and the faithfulness assumption.} 

However, existing constrained-based structural learning algorithms can be quite aggressive in edge removal: 
{if a type II error is made such that two connected covariates $X$ and $Y$ are thought to be conditionally independent given some $\tf{Z}$,
the edge between $X$ and $Y$ is removed mistakenly,
and useful information about important features and possible causal pathways get lost during this edge removal process.}
\textcolor{black}{Especially in the case of categorical variables and relatively small sample sizes, the conditional independence tests used by structural learning algorithms can have high type II error rates that lead to many false edge-removals and a severe information loss.}
As illustrated in the PHIA data analysis, very few features are connected by using the constrained-based structural learning algorithms. Results and discussions in detail can be found later in Section~\ref{section:Results}.

\textcolor{black}{In literature dealing with the unreliable conditional independence tests, most literature focuses on their negative effects in the orientation procedure but much less on those in the skeleton learning procedure, with a few exceptions. \citet{bromberg2009improving} proposed a method to resolve the inconsistencies in the conditional independence tests in the skeleton learning procedure. Their idea is to deduce a ``preference" score on the conditional independence test results from a certain set of axioms, such as Pearl's axioms \citep{dawid1979conditional, pearl1988probabilistic}, which all true conditional independence relationships should follow. These axioms can be seen as integrity constraints that can avoid certain inconsistent test outcomes.
However, the computational complexity of the algorithm proposed by \citet{bromberg2009improving} significantly increases as the number of vertices increases. Thus, the algorithm cannot be employed on the MPHIA dataset for which the graph is not sparse enough.
}

In this article, we propose a new causal structural learning algorithm that aims to preserve more information on important features and potential causal pathways. We apply the proposed algorithm on the Malawi PHIA \citep{ministry2017malawi} data set, which we refer to as MPHIA, and obtain interesting findings related to Tri90 pathways. We further compare and validate the proposed algorithm with some classical structural learning algorithms using information criteria and simulations.

The remaining part of the paper is organized as follows. In Section~\ref{section:Preliminaries}, we provide definitions for important concepts in causal structural learning, such as d-separation, and also a basic summary for the graphical notations used in the paper.
In Section~\ref{section:Motivation}, we use a simple example to illustrate the problem of aggressive edge removal of existing structural learning algorithms. In Sections~\ref{section:MethodEquation} $\sim$ \ref{section:Overall}, we propose a new causal structural learning algorithm to overcome the aggressive edge removal issue.
In Section~\ref{section:Results}, we apply the proposed algorithm on the MPHIA data set, compare the results obtained from the proposed algorithm with those of the existing algorithms in Section~\ref{section:Validation}, and discuss the discovered Tri90 pathways in detail in Section~\ref{section:Discussions}.
In Section~\ref{section:Simulations},
we compare the numerical performance of the proposed algorithm with classical structural learning algorithms in simulation studies.
Section~\ref{section:Conclusions} provides a summary and discussion for the paper.
To save space, technical details, additional numerical results, and the relevant codebooks are provided in the Supplement.

\section{Method}
\label{section:Method}

In this section, we propose our causal structural learning algorithm.
We first provide definitions for important concepts in causal structural learning, such as d-separation, and also a basic summary for the graphical notations used in the paper in Section~\ref{section:Preliminaries}.
We then discuss the problem of aggressive edge removal of existing structural learning algorithms with a simplified example in Section~\ref{section:Motivation}.
To deal with this problem, we propose our new algorithm and provide a high-level overview of the algorithm in Section~\ref{section:MethodEquation}.
The proposed algorithm consists of two main steps, the forward step and the maximization step, which are explained in detail in Sections~\ref{section:Step1} and \ref{section:Step2}, respectively.
Section~\ref{section:Overall} provides the orientation procedure of our proposed algorithm and also summarizes the proposed method.

\subsection{Preliminaries}
\label{section:Preliminaries}


\textcolor{black}{
Suppose $\mathcal{G}$ is a directed acyclic graphical (DAG) model which represents the joint probability distribution over the vertex set $\tf{V}$ with directed edges and no directed loop. Each vertex in the graph represents a variable. We use variables $X$, $Y$, etc., to refer to the variables corresponding to the vertices $X$, $Y$, etc., and use the edge $X \rightarrow Y$ or $Y \leftarrow X$ to refer to the directed edge from $X$ to $Y$, where $X$ is a parent vertex of $Y$, and $Y$ is a child of $X$.
$X - Y$ denotes an undirected edge that could be either $X \rightarrow Y$ or $Y \rightarrow X$. We use a path to refer to an acyclic sequence of adjacent vertices and a causal path from $X$ to $Y$ to refer to a path that all arrows are pointing away from $X$ and into $Y$. If there is a causal path from $X$ to $Y$, we say that $X$ is an ancestor of $Y$ and that $Y$ is a descendant of $X$.}

\textcolor{black}{Next, we provide the formal definition of d-separation \citep{geiger1990logic, geiger1990d}.
A collider is a vertex on a path with two incoming arrows.
More specifically, a vertex $Z$ is a collider (v-structure) on a path $U$ if and only if the path $U$ contains a subpath $X \rightarrow Z \leftarrow Y$. 
For vertices $X$, $Y$ and a vertex set  $\tf{Z}$ which does not contain $X$ and $Y$, $X$ is d-connected to $Y$ given $\tf{Z}$ if and only if there is an acyclic path $U$ between $X$ and $Y$ such that every collider on $U$ is either a member of $\tf{Z}$ or an ancestor of a member of $\tf{Z}$, and no non-collider on $U$ is in $\tf{Z}$.
$X$ is d-separated from $Y$ given $\tf{Z}$ if and only if $X$ is not d-connected to $Y$ given $\tf{Z}$. As a simple illustration, suppose that $Z$ is the only vertex in $\tf{Z}$ and $U$. $X$ and $Y$ are d-connected given $Z$ for $X \rightarrow Z \leftarrow Y$; $X$ and $Y$ are d-separated given $Z$ for $X \rightarrow Z \rightarrow Y$, $X \leftarrow Z \leftarrow Y$, and $X \leftarrow Z \rightarrow Y$.} 


\textcolor{black}{Moreover, a set of variables $\tf{V}$ is causally sufficient if and only if no variable outside $\tf{V}$ is a direct cause of more than one variable in $\tf{V}$. For a causally sufficient set of variables $\tf{V}$ with probability distribution $P(\tf{V})$, the Markov condition assumes that the d-separation in the DAG $\mathcal{G}$ implies conditional independence in $P(\tf{V})$, i.e. if $X$ is d-separated from $Y$ by $\tf{Z}$ in $\mathcal{G}$, then $X$ is independent of $Y$ conditional on $\tf{Z}$ in $P(\tf{V})$; and the faithfulness assumption assumes that every conditional independence relationship in $P(\tf{V})$ is entailed by the d-separation relationship for the causal DAG $\mathcal{G}$, i.e. if $X$ is independent of $Y$ conditional on $\tf{Z}$, then $X$ is d-separated from $Y$ by $\tf{Z}$.
Therefore, the d-separation in the graph is equivalent to conditional independence in the distribution under the Markov condition and the faithfulness assumption. See \citet{spirtes2010introduction} for a more detailed introduction and discussion. In this paper, we always assume the causal sufficiency, the Markov condition, and the faithfulness assumption.}

\subsection{A Motivation Example}
\label{section:Motivation}


In the MPHIA data set, classical constrained-based structural learning algorithms discover very few important features for Tri90, as shown later in Section~\ref{section:Results}.
In this section, we explain the main cause of the problem and illustrate the motivation of our method with a simplified example from the MPHIA data. 
A complete analysis of the MPHIA data can be seen in Section~\ref{section:Results}.

\begin{example}
    \label{ex:ex1}
    \textcolor{black}{Suppose our graph contains only three vertices $X$, $Y$ and $Z$, and $X$ is Tri90Aware, the indicator variable for awareness of the HIV positive status, which is one of the variables in which we are mainly interested in our Tri90 goal study. 
    We want to check whether $Y$ (AlcoholFrequency, an ordinal variable for alcohol drinking frequency) and $Z$ (WealthQuintile, an ordinal variable for the wealthiness) are neighbors of $X$ (Tri90Aware). 
    That is to say, we are interested in whether the wealthiness and alcohol drinking frequency have direct causal relationships with the awareness of the HIV positive status.}
    We have the following four (conditional) independence test results among $X$, $Y$, and $Z$ for a confidence level $\alpha=0.05$ in the MPHIA dataset of males who are included in the Tri90 study.
    \begin{equation}
    \label{eq:testing}
    \begin{aligned}
    & X\perp Y \text{ is rejected, and } X\perp Z \text{ is rejected}, \\
    & X\perp Y| Z \text{ is not rejected, and } X\perp Z| Y \text{ is not rejected}.
    \end{aligned}
    \end{equation}
    
    \textcolor{black}{Since both $Y$ (AlcoholFrequency) and $Z$ (WealthQuintile) are not independent with $X$ (Tri90Aware) marginally from the testing results,
    they should be connected to $X$ (Tri90Aware) either directly or indirectly, according to the Markov condition.
    However, based on the faithfulness assumption,
    neither $Y$ (AlcoholFrequency) nor $Z$ (WealthQuintile) should be a neighbor of $X$ (Tri90Aware) since $X$ (Tri90Aware) is conditionally independent with $Y$ (AlcoholFrequency) given $Z$ (WealthQuintile) and $X$ (Tri90Aware) is conditionally independent with $Z$ (WealthQuintile) given $Y$ (AlcoholFrequency).}
    Those four testing results lead to contradicted conclusions under the Markov condition and the faithfulness assumption.
    Existing structural learning algorithms, such as the PC-stable algorithm, remove both the edge between Tri90Aware and AlcoholFrequency ($X - Y$) and the edge between Tri90Aware and WealthQuintile ($X - Z$) from the conditional independence testing results and conclude that there are no neighbors of $X$ (Tri90Aware), which might be too strict in detecting edges.

    
\end{example}

\textcolor{black}{Many covariates in the MPHIA data are categorical. When used as the conditional set, those categorical covariates lead to relatively high type II error rates for the conditional independence tests. In addition, triples $(X,Y,Z)$ with contradictory/inconsistent testing results such as Equation~(\ref{eq:testing}) are quite common in the MPHIA data. Such contradiction/inconsistency also leads to aggressive edge-removal for the constrained-based causal structural learning algorithms.}
To solve the false edge-removal issue, we propose a new graphical structural learning algorithm,
and we illustrate how the new algorithm successfully finds more edges later in Example~\ref{ex:ex2}.

\subsection{Method}
\label{section:MethodEquation}

In a directed acyclic graph (DAG) $\mathcal{G}$ with vertex set $\tf{V}$ and $X \in \tf{V}$.
Suppose $\tf{N}_X=(N_1,N_2,\cdots,N_q)$ is the parents and children set of $X$.
Under the faithfulness assumption and the Markov condition, respectively,
we have
\begin{eqnarray}
\label{eq:1}
&&X\not\perp N_i | \tf{S} \text{ for any } \tf{S}\subseteq(\tf{N}\backslash\{N_i\}), i=1,\cdots,q,\\
\label{eq:2}
&&X\perp M_i | \tf{S}_i \text{ for some } \tf{S}_i\subseteq\tf{N}_X, i=1,\cdots,r,
\end{eqnarray}
where
$
\tf{M} = \{M_1,M_2,\cdots,M_r\}
= \tf{V} \backslash (\tf{N}_X \cup \{X\})
$
is the set of vertices not connected to $X$.
Furthermore, under the Markov condition and the faithfulness assumption,
suppose the variable set, $\tf{N}$ satisfies the conditions \eqref{eq:1} and \eqref{eq:2},
then it is easy to verify that $\tf{N}$ is the set of parents and children of $X$ and thus $\tf{N} = \tf{N}_X$. 

For each vertex $X$, we want to get the best subset of $\tf{V}$ that fulfills the conditions \eqref{eq:1} and \eqref{eq:2}.
Our procedure consists of two steps: a forward step and a maximization step.
The forward step finds all sets $\tf{N}$ that satisfy \eqref{eq:1} for vertex $X$ and will be illustrated in Algorithm~\ref{alg:step1} in Section~\ref{section:Step1}.
The maximization step picks the best set that fulfills \eqref{eq:2} among those sets found by Algorithm~\ref{alg:step1} and will be provided in Section~\ref{section:Step2}.


Before deriving the details of the algorithms, let us come back to Example~\ref{ex:ex1} and see how algorithms based on \eqref{eq:1} and \eqref{eq:2} can solve the problem of aggressive edge removal that presents in the existing classical structural learning algorithms.

\begin{example}
    \label{ex:ex2}
    \textcolor{black}{Same as Example~\ref{ex:ex1}, suppose our graph only contains three vertices $X$ (Tri90Aware), $Y$ (AlcoholFrequency), and $Z$ (WealthQuintile).}
    Further, assume that we have the same (conditional) independence test results as in Example~\ref{ex:ex1}.
    Although the four testing results in \eqref{eq:testing} are incompatible under the Markov condition and the faithfulness assumption,
    it is possible to have a valid graphical structure with only three of them.
    For example,
    \begin{equation}
    \label{eq:ex2(1)}
    \begin{tikzpicture}
	\begin{pgfonlayer}{nodelayer}
		\node [style=simple] (0) at (-6, -1.25) {$Z$};
		\node [style=simple] (1) at (-7.75, 0.75) {$X$};
		\node [style=simple] (2) at (-9.5, -1.25) {$Y$};
		\node [style=none, label={above:$X\not\perp Y, X\not\perp Z, X\perp Y|Z$}] (3) at (0, 3.75) {};
		\node [style=simple] (4) at (1.75, -1.25) {$Z$};
		\node [style=simple] (5) at (0, 0.75) {$X$};
		\node [style=simple] (6) at (-1.75, -1.25) {$Y$};
		\node [style=none] (7) at (0, 1.25) {};
		\node [style=none] (8) at (-4, 0) {};
		\node [style=none] (9) at (-4, 0) {or};
		\node [style=none] (10) at (0, 2) {};
		\node [style=simple] (11) at (9.5, -1.25) {$Z$};
		\node [style=simple] (12) at (7.75, 0.75) {$X$};
		\node [style=simple] (13) at (6, -1.25) {$Y$};
		\node [style=none] (14) at (3.75, 0) {};
		\node [style=none] (15) at (3.75, 0) {or};
	\end{pgfonlayer}
	\begin{pgfonlayer}{edgelayer}
		\draw [style=arrow] (1) to (0);
		\draw [style=arrow] (0) to (2);
		\draw [style=arrow] (4) to (6);
		\draw [style=arrow] (3.center) to (10.center);
		\draw [style=arrow] (4) to (5);
		\draw [style=arrow] (11) to (12);
		\draw [style=arrow] (13) to (11);
	\end{pgfonlayer}
\end{tikzpicture}

    \end{equation}
    and
    \begin{equation}
    \label{eq:ex2(2)}
    \begin{tikzpicture}
	\begin{pgfonlayer}{nodelayer}
		\node [style=simple] (0) at (-6, -1.25) {$Z$};
		\node [style=simple] (1) at (-7.75, 0.75) {$X$};
		\node [style=simple] (2) at (-9.5, -1.25) {$Y$};
		\node [style=none, label={above:$X\not\perp Y, X\not\perp Z, X\perp Z|Y$}] (3) at (0, 3.75) {};
		\node [style=simple] (4) at (1.75, -1.25) {$Z$};
		\node [style=simple] (5) at (0, 0.75) {$X$};
		\node [style=simple] (6) at (-1.75, -1.25) {$Y$};
		\node [style=none] (7) at (0, 1.25) {};
		\node [style=none] (8) at (-4, 0) {};
		\node [style=none] (9) at (-4, 0) {or};
		\node [style=none] (10) at (0, 2) {};
		\node [style=simple] (11) at (9.5, -1.25) {$Z$};
		\node [style=simple] (12) at (7.75, 0.75) {$X$};
		\node [style=simple] (13) at (6, -1.25) {$Y$};
		\node [style=none] (14) at (3.75, 0) {};
		\node [style=none] (15) at (3.75, 0) {or};
	\end{pgfonlayer}
	\begin{pgfonlayer}{edgelayer}
		\draw [style=arrow] (3.center) to (10.center);
		\draw [style=arrow] (1) to (2);
		\draw [style=arrow] (2) to (0);
		\draw [style=arrow] (6) to (5);
		\draw [style=arrow] (6) to (4);
		\draw [style=arrow] (11) to (13);
		\draw [style=arrow] (13) to (12);
	\end{pgfonlayer}
\end{tikzpicture}

    \end{equation}
    
    \textcolor{black}{Note that the neighborhood of $X$ (Tri90Aware) satisfies \eqref{eq:1} and \eqref{eq:2} for $X$.
    In the situations of \eqref{eq:ex2(1)}, the neighbor of $X$ (Tri90Aware) is $Z$ (WealthQuintile);
    and in the situations of \eqref{eq:ex2(2)}, the neighbor of $X$ (Tri90Aware) is $Y$ (AlcoholFrequency).
    However, $\{Y,Z\}$ does not satisfy \eqref{eq:1} and is not the neighborhood of $X$.
    Algorithm~\ref{alg:step1} finds both $\{Z\}$ and $\{Y\}$ as potential neighborhoods for $X$ (Tri90Aware) since both of them satisfy \eqref{eq:1} for $X$.
    Algorithm~\ref{alg:step2} compares between $\{Z\}$ and $\{Y\}$ and chooses the set which fulfills \eqref{eq:2} better as the neighborhood of $X$.
    In sum, the new algorithm chooses either $\{Z\}$ or $\{Y\}$ as the neighborhood of $X$, and concludes that either AlcoholFrequency ($Y$) or WealthQuintile ($Z$) is the neighbor of Tri90Aware ($X$) but not both of them.}
    So, the proposed algorithm is less aggressive in edge removal than the existing classical structural learning algorithms such as the PC-stable algorithm as discussed in Example~\ref{ex:ex1}.

\end{example}

\subsection{Forward Step}
\label{section:Step1}

Before stating Algorithm~\ref{alg:step1}, we first provide some useful definitions and their properties. Let $p$ be the total number of vertices.
For each vertex $X$, let $\tf{T} = \tf{V}\backslash \{X\} = \{T_i, i=1,2,\cdots,p-1\}$
and vertices are considered/added sequentially in the order of $T_1,T_2,\cdots,T_{p-1}$ to form a candidate neighborhood of $X$. That is to say, for any already formed non-empty candidate neighborhood $\tf{S} = \{T_{s_1},\cdots, T_{s_q}\} \subseteq \tf{T}$, where $1\leq s_1 < s_2 < \cdots < s_q \leq p-1$, we consider whether an additional vertex from $\tf{L}_X(\tf{S}):=\{T_{{s_q}+1},T_{s_q+2},\cdots,T_{p-1}\}$ can be added into the variable set, $\tf{S}$. We also define $\tf{L}_X(\emptyset)=\tf{T}$, which means that we need to consider all the vertices in $\tf{T}$ when $\tf{S}$ starts from an empty set.
Furthermore, let $\tf{C}_{X}(\tf{S})$ be the vertices in $\tf{L}_X(\tf{S})$ that can be added into $\tf{S}$ while still satisfying equation \eqref{eq:1}. That is to say,

\begin{equation}
\tf{C}_{X}(\tf{S}) := \{
T \in \tf{L}_X(\tf{S}) | \tf{N} = (\tf{S} \cup \{T\}) \text{ satisfies equation \eqref{eq:1}}
\}.
\end{equation}

Proposition~\ref{prop:cxs} establishes some useful properties of $\tf{C}_X(\tf{S})$, which are used in Algorithm~\ref{alg:step1} to facilitate the calculation of $\tf{C}_X(\tf{S})$.

\begin{proposition}
    \label{prop:cxs}

    $\tf{C}_X(\tf{S})$ has the following properties.
    \begin{enumerate}[topsep=0pt,itemsep=-1ex,partopsep=0ex,parsep=0ex]
        \item Let $\emptyset$ denote the empty set. If $\tf{S}=\emptyset$, then 
        \begin{equation}
        \label{eq:initialC}
        \tf{C}_X(\emptyset) = \{C|C\not\perp X\}.
        \end{equation}

        \item If $\tf{S}$ does not satisfy \eqref{eq:1}, then
        \begin{equation}
        \label{eq:emptyC}
        \tf{C}_X(\tf{S}) = \emptyset.
        \end{equation}

        \item If $\tf{S}_1\subseteq\tf{S}$, then
        \begin{equation}
        \label{eq:inclusion}
        \tf{C}_X(\tf{S}) \subseteq \tf{C}_X(\tf{S}_1).
        \end{equation}

        \item Let $\tf{S} = \{S_1,S_2,\cdots,S_n\}$, $n\geq 1$,
        $\tf{C}_X^{*}(\tf{S}) := \bigcap_{i=1}^n \tf{C}_X(\tf{S}_{-i})$,
        and $\tf{S}_{-i} = \tf{S} \backslash \{S_i\}$, for $i=1,2,\cdots,n$. It follows that
        \begin{equation}
        \label{eq:upperbound}
        \tf{C}_X(\tf{S}) \subseteq \tf{C}_X^{*}(\tf{S}).
        \end{equation}

        \item Let $\tf{S} = \{S_1,S_2,\cdots,S_n\}$, $n\geq 1$. If
        $\tf{S}$ satisfies \eqref{eq:1}, then
        \begin{equation}
        \label{eq:incremental_test}
        \tf{C}_X(\tf{S}) =
        \{C \in \tf{C}_X^{*}(\tf{S}) \cap \tf{L}_X(\tf{S}) :
        C\not\perp X|\tf{S},
        S_i\not\perp X| (\tf{S}_{-i} \cup \{C\}), i=1,\cdots,n
        \}.
        \end{equation}

    \end{enumerate}

\end{proposition}

The properties of $\tf{C}_X(\tf{S})$ in Proposition~\ref{prop:cxs} can be shown by using its definition,
and the proof can be found in Supplement~\ref{section:proof_cxs}.
Proposition~\ref{prop:cxs} establishes a recursive structure for $\tf{C}_X(\tf{S})$.
\eqref{eq:initialC} shows that $\tf{C}_X(\tf{S})$ consists of all vertices that are marginally correlated with $X$ when $\tf{S}$ is an empty set.
\eqref{eq:upperbound} states that $\tf{C}_X^{*}(\tf{S})$ is an upper-bound (concerning the partial order of inclusion) for $\tf{C}_X(\tf{S})$ when $\tf{S}$ is non-empty.
Furthermore, \eqref{eq:incremental_test} tells us
that we only need to examine the conditional independence relationship with conditional set of size $|\tf{S}|$ to get $\tf{C}_X(\tf{S})$ from $\tf{C}_X^{*}(\tf{S}) \cap \tf{L}_X(\tf{S})$ for non-empty $\tf{S}$.
\eqref{eq:incremental_test} is used in Algorithm~\ref{alg:step1} to reduce the number of conditional independence tests and speed up the finding process.
We can show that using \eqref{eq:incremental_test},
no conditional independence test is repeated in Algorithm~\ref{alg:step1}. Indeed,
let $s_{\max} = \max\{s_0,s_1,\cdots,s_q\}$, $S_{\max} = T_{s_{\max}}$, 
$\tf{s}_{-\max} = \{s_0,s_1,\cdots,s_q\}\backslash\{s_{\max}\}$,
and $\tf{S}_{-\max} = \{T_s| s\in\tf{s}_{-\max}\}$, 
then the conditional independence test of $X$ and $T_{s_0}$ given $\tf{S}=\{T_{s_1},T_{s_2}\cdots,T_{s_q}\}$ can only happen when we check whether $S_{\max}$ can be added into candidate set $\tf{S}_{-\max}$.

\begin{singlespace}
\begin{algorithm}[h]
    \caption{Forward Step: Find all potential parent and children sets of $X$ that satisfy  \eqref{eq:1}. The algorithm sequentially adds the candidate set satisfying \eqref{eq:1} into the preliminary result set, $\mathcal{R}$.}
    \label{alg:step1}
    \begin{algorithmic}
        \REQUIRE{
            (1) a vertex set $\textbf{V}$,
            (2) a vertex ${X}$ in $\tf{V}$, and
            (3) size $\alpha$ for conditional independence tests.
        }
        \ENSURE{The set $\mathcal{N}_X$ including all possible $\tf{N}$ that (approximately) satisfies equation \eqref{eq:1}.}
        \STATE{Let $\mathcal{S} = \{\emptyset\}$ and $\mathcal{R} = \emptyset$.} 
        \WHILE{$\mathcal{S}\neq\emptyset$}
        \STATE{Calculate $\tf{C}_X(\tf{S})$ for each $\tf{S}\in\mathcal{S}$ of cardinal from low to high as follows:}
        \IF{$\tf{S} = \emptyset$}
        \STATE{$\tf{C}_X(\emptyset) = \{C|C\not\perp X \text{ under size } \alpha\}$.}
        \ELSE
        \IF{$|\tf{S}| > M_{\text{CI}}$}
        \STATE{Set $\tf{C}_X(\tf{S}) = \tf{C}_X^{*}(\tf{S}) \cap \tf{L}_X(\tf{S})$. }
        \ELSE
        \STATE{Calculate $\tf{C}_X(\tf{S})$ from equation \eqref{eq:incremental_test}.}
        \ENDIF
        \ENDIF

        \STATE{Delete $\tf{S}$ from $\mathcal{S}$.}
        \IF{$\tf{C}_X(\tf{S})=\emptyset$}
        \STATE{Add $\tf{S}$ into the preliminary result set, $\mathcal{R}$.}
        \ELSE
        \STATE{Add every $\tf{S}\cup\{C\}$ for $C$ in $\tf{C}_X(\tf{S})$ into $\mathcal{S}$.}
        \ENDIF
        \ENDWHILE
        \RETURN{$\mathcal{N}_X = \{\tf{R}|\tf{R}\in\mathcal{R}, \text{ and \tf{R} is not a proper subset of any other sets in } \mathcal{R}\}$.}
    \end{algorithmic}
\end{algorithm}
\end{singlespace}

\noindent\tf{Remark 1}: 
Note that the final output of Algorithm~\ref{alg:step1}, $\mathcal{N}_X$, only contains the candidate sets which are not proper subsets of other candidate sets for the following reasons. If $\tf{N}_X$ is a candidate set that satisfies \eqref{eq:1} for $X$, then all proper subsets of $\tf{N}_X$ satisfy \eqref{eq:1} for $X$, but no proper subset of $\tf{N}_X$ will satisfy \eqref{eq:2}. This is because
for any $\tf{N}_1\subsetneq \tf{N}_X$, take $M\in\tf{N}_X\backslash\tf{N}_1$, and we have $X\not\perp M|\tf{S}$ for any $\tf{S}\subseteq\tf{N}_1\subsetneq\tf{N}_X$ since $\tf{N}_X$ satisfies \eqref{eq:1}. Hence $\tf{N}_1$ does not satisfy \eqref{eq:2}.

\noindent\tf{Remark 2}: In the calculation of $\tf{C}_X(\tf{S})$ and $\tf{C}_X^{*}(\tf{S})$, the size of conditional sets in the conditional independence test is not restricted.
However, conditional independence tests with large conditional sets are not accurate, and the number of all possible conditional independence tests also grows exponentially with the size of conditional sets.
Hence we set the upper-bound of sizes of conditional sets in Algorithm~\ref{alg:step1},
which is also discussed by \citet{tsamardinos2006mmhc}
and is commonly implemented in causal structural learning software such as \citet{scutari2010bnlearn}.
In addition, as shown by \eqref{eq:incremental_test}, if we set the upper-bound of sizes of conditional sets to be $M_{\text{CI}}$, then we can approximate $\tf{C}_X(\tf{S})$ by
\begin{equation}
\tf{C}_X(\tf{S}) \approx
\tf{C}_X^{*}(\tf{S}) \cap \tf{L}_X(\tf{S}), \text{ if } |\tf{S}| > M_{\text{CI}}.
\end{equation}

\textcolor{black}{\noindent{\bf Remark 3}:
$M_{\text{CI}}$ is a tuning parameter in Algorithm 1.
Theoretically, one should choose $M_{\text{CI}}$ that is no less than the largest degree of vertices in the graph. However, there are concerns about using a large $M_{\text{CI}}$, and the true degrees are also unknown in practice.
First, a large $M_{\text{CI}}$ corresponds to big conditional sets, which will increase the computing cost. Second, when the sample size is not large enough, a large $M_{\text{CI}}$ leads to less reliable results due to the increased type II errors. 
Generally, it is recommended to set $M_{\text{CI}}=3$ when the true graph is expected to be sparse or moderately sparse \citep{yan2020effective}. We also carry on a simulation study on different choices of $M_{\text{CI}}$, and more details and discussions can be seen in Supplement~S.5.1.
}

\subsection{Maximization Step}
\label{section:Step2}

To choose the best neighbor set from the set of candidate sets $\mathcal{N}_X$, we check how well each candidate $\tf{N} \in \mathcal{N}_X$ satisfies equation \eqref{eq:2}.
Define
\begin{equation}
\label{eq:snxy}
S_N(X, Y) := \max_{S\subseteq N} \CI(X,Y|\tf{S}),
\end{equation}
where $\CI(X,Y|\tf{S})$ is the $p$-value of some chosen conditional independence test for $X$ and $Y$ given $\tf{S}$.
\textcolor{black}{
To see why we use $\max$ in \eqref{eq:snxy}, note that conditional independence relationship implies big $p$-values, and $S_N(X,Y)$ measures whether any subset of $\tf{N}$ makes $X$ and $Y$ conditional independent.
The maximum in \eqref{eq:snxy} is from the fact that \eqref{eq:2} only requires one $\tf{S}$ that makes $X$ and $Y$ conditionally independent given $\tf{S}$ as suggested by the large $p$-value.
The idea of using conditional independence set of the largest $p$-value in the causal structural learning algorithm can also be found in \citet{ramsey2016improving}.}

Define
\begin{equation}
\label{eq:qxn}
Q_X(\tf{N}) := \min_{M_i\in \tf{V} \backslash (\tf{N}\cup\{X\})} S_N(M_i,X),
\end{equation}
which measures how well it is for subsets of $\tf{N}$ to ``separate" $X$ from any vertices not in $\tf{N}\cup\{X\}$. \textcolor{black}{Here we are concerned about whether there is any $M_i\in \tf{V} \backslash (\tf{N}\cup\{X\})$ violates the conditional independence {between $M_i$ and $X$ given $\tf{N}$}, and hence the minimum of $p$-values across $M_i$'s is used. Equations~\eqref{eq:snxy} and \eqref{eq:qxn} together can be seen as a minimax procedure. If $\tf{N}$ satisfies Equation \eqref{eq:2}, then for any $M_i$ not in the neighborhood of $X$, there should exist $\tf{S}\subseteq\tf{N}$ such that $M_i\perp X|\tf{S}$ leading to a large value of $S_N(M_i, X)$. If $Q_X(\tf{N})$ is large, then for any $M_i$ not in the neighborhood of $X$, $S_N(M_i, X)$ is large and it is likely that $M_i\perp X|\tf{S}$. So $\tf{N}$ with larger $Q_X(\tf{N})$ is likely to be the true neighbor set of $X$.} Furthermore, we have the following useful properties of $S_N(X, Y)$ in Proposition~\ref{prop:snxy}, which are used in 
Algorithm~\ref{alg:step2} to facilitate the calculation of $S_N(X, Y)$.

\begin{proposition}
    \label{prop:snxy}

    $S_N(X, Y)$ has the following properties.
    \begin{enumerate}[topsep=0pt,itemsep=-1ex,partopsep=0ex,parsep=0ex]
        \item Let $\emptyset$ be the empty set. Then
        \begin{equation}
        \label{eq:initialS}
        S_{\emptyset}(X, Y) = \CI(X,Y,\emptyset).
        \end{equation}

        \item If  $\tf{N}_1\subseteq\tf{N}$, then
        \begin{equation}
        \label{eq:inclusionS}
        S_N(X, Y) \geq S_{N_1}(X, Y).
        \end{equation}

        \item  Let $\tf{N} = \{N_1,N_2,\cdots,N_n\}$, $n\geq 1$,
        $S_N^{*}(X, Y) := \max_{i=1}^n S_{N_{-i}}(X, Y)$,
        and $\tf{N}_{-i} = \tf{N} \backslash \{N_i\}$, $i=1,2,\cdots,n$. Then
        \begin{equation}
        \label{eq:lowerboundS}
        S_N(X, Y) \geq S_N^{*}(X, Y).
        \end{equation}

        \item If $\tf{N}\neq\emptyset$, then
        \begin{equation}
        \label{eq:incremental_calcS}
        S_N(X, Y) = \max\{S_N^{*}(X, Y), \CI(X,Y|\tf{N})\}.
        \end{equation}

    \end{enumerate}

\end{proposition}

The properties of $S_N(X, Y)$ in Proposition~\ref{prop:snxy} can be shown by using its definition,
and the proof can be found in Supplement~\ref{section:proof_snxy}.
Similar to Proposition~\ref{prop:cxs}, Proposition~\ref{prop:snxy} shows a recursive structure in $S_N(X, Y)$.
\eqref{eq:initialS} shows how to calculate $S_N(X, Y)$ for an empty set.
\eqref{eq:lowerboundS} states that $S_N^{*}(X, Y)$ is a lower-bound for $S_N(X, Y)$ for a non-empty set $\tf{N}$.
Furthermore, \eqref{eq:incremental_calcS} tells us
that we only need to calculate the $p$-value of conditional independence test $ \CI(X,Y|\tf{N})$ to get $S_N(X, Y)$ from $S_N^{*}(X, Y)$ when $\tf{N}$ is non-empty.
\eqref{eq:incremental_calcS} is used in
Algorithm~\ref{alg:step2} to reduce the number of conditional independence tests and speed up the searching process.
It is easy to prove that using \eqref{eq:incremental_calcS},
no conditional independence test is repeated in 
Algorithm~\ref{alg:step2} because the conditional independence test of $X$ and $Y$ given $\tf{N}$ only happens when we calculate $S_{N}(X,Y)$.

\noindent\tf{Remark}:
Similar to Algorithm~\ref{alg:step1}, we propose to set upper-bound for sizes of conditional sets to be $M_{\text{CI}}$,
then $S_N(X, Y)$ can be approximated by
\begin{equation}
\label{eq:appox_S}
S_N(X, Y) \approx \max_{S\subseteq N, |S| \leq M_{\text{CI}}} \CI(X,Y|\tf{S}).
\end{equation}
Furthermore, from \eqref{eq:incremental_calcS}, we can approximate $S_N(X, Y)$ by
\begin{equation}
\label{eq:incremental_calcS_approx}
S_N(X, Y) \approx
S_N^{*}(X, Y), \text{ if } |\tf{N}| > M_{\text{CI}}.
\end{equation}

Based on \eqref{eq:incremental_calcS} and \eqref{eq:incremental_calcS_approx}, we have Algorithm~\ref{alg:step2} which selects the $\tf{N}$ with the largest $Q_X(\tf{N})$ to be the neighbor set of $X$.

\begin{singlespace}
\begin{algorithm}[!h]
    \caption{Maximization Step: Find $\tf{N} \in \mathcal{N}_X$ with the largest $Q_X(\tf{N})$.}
    \label{alg:step2}
    \begin{algorithmic}
        \REQUIRE{
            (1) a vertex set $\textbf{V}$,
            (2) a vertex ${X}$ in $\tf{V}$, and
            (3) candidate neighbor set $\mathcal{N}_X$ of $X$.}
        \ENSURE{$\tf{N} \in \mathcal{N}_X$ with the largest $Q_X(\tf{N})$.}
        \STATE{Let $r = 0$ and $\tf{N}_X = \emptyset$.}
        \FOR{$\tf{N} \in \mathcal{N}_X$}
        \STATE{Let $\tf{M} = \tf{V} \backslash (\tf{N}\cup\{X\})$.}
        \STATE{Set $Q_X(\tf{N}) = \infty$.}
        \FOR{each vertex $M_i \in \tf{M}$}
        \IF{$\tf{N} = \emptyset$}
        \STATE{$S_N(M_i, X) = \CI(M_i,X,\emptyset)$.}
        \ELSE
        \STATE{Calculate $S_N(M_i, X)$ from equation \eqref{eq:incremental_calcS} or approximate $S_N(M_i, X)$ using \eqref{eq:incremental_calcS_approx}.}
        \ENDIF
        \IF{$S_N(M_i, X) \leq r$}
        \STATE{Continue the outer loop for next $\tf{N} \in \mathcal{N}_X$.}
        \ENDIF
        \STATE{Set $Q_X(\tf{N}) = \min\{Q_X(\tf{N}), S_N(M_i, X)\}$.}
        \ENDFOR
        \IF{$Q_X(\tf{N}) > r $}
        \STATE{Set $\tf{N}_X = \tf{N}$.}
        \STATE{Set $r = Q_X(\tf{N})$.}
        \ENDIF
        \ENDFOR
        \RETURN{$\tf{N} = \tf{N}_X$.}
    \end{algorithmic}
\end{algorithm}
\end{singlespace}

\subsection{Overall Structural Learning with Orientations}
\label{section:Overall}

With Algorithms~\ref{alg:step1} and \ref{alg:step2}, we can learn the neighbor set of every vertex $X$ in the DAG $\mathcal{G}$ and hence the skeleton of the DAG.
We can further orient edges according to the conditional independence relationship.
The overall structural learning algorithm is summarized in Algorithm~\ref{alg:structure_learn}. The orientation procedure is quite similar to those of classic constraint-based causal structural learning algorithms, such as the PC algorithm \citep{spirtes1991algorithm}. We also incorporate some prior knowledge into the orientations.
For example, we know covariates such as AgeGroup cannot be affected by other covariates such as Education, so if there is an edge between AgeGroup and Education, then we orient the edge as AgeGroup to Education.
If there are multiple d-separation sets $S(X, Y)$ for a non-adjacent pair $(X, Y)$, we shall use the d-separation set $S(X,Y)$ with the largest $p$-value to make the orientation results stable.
For $Z$ adjacent to both $X$ and $Y$, we call $X-Z-Y$ a v-structure and make the orientation $X\rightarrow Z\leftarrow Y$ if $Z\not\in S(X,Y)$.
Finally, note that there could be conflicting v-structures.
For example, if there is a chain $X-Z-Y-W$, and both $X-Z-Y$ and $Z-Y-W$ are v-structures, then the edge $Z-Y$ should be oriented as $Z\leftarrow Y$ from the v-structure of $X-Z-Y$ but $Z\rightarrow Y$ from the v-structure of $Z-Y-W$.
In Algorithm~\ref{alg:structure_learn}, we resolve the conflict between v-structures by comparing the $p$-value of the v-structure.
Going back to the previous example, if $S(X,Y)$ has a larger $p$-value than $S(Z,W)$, then we orient $Z-Y$ as $Z\leftarrow Y$ following v-structure $X-Z-Y$ and vice-versa.

\begin{singlespace}
\begin{algorithm}[t]
    \caption{Structural Learning Algorithm}
    \label{alg:structure_learn}
    \begin{algorithmic}
        \REQUIRE{
            (1) A vertex set $\textbf{V}$, and
            (2) size $\alpha$ for conditional independence tests.
        }
        \ENSURE{CPDAG $\mathcal{G}$.}
        \FOR{every vertex $X \in \tf{V}$}
        \STATE{Calculate the neighbor set $\tf{N}_X$ using Algorithms~\ref{alg:step1} and \ref{alg:step2}.}
        \ENDFOR
        \STATE{Start from a complete undirected graph $\mathcal{G}$ with the vertex set $\tf{V}$.}
        \FOR{every pair of vertices $(X,Y)$ connected in $\mathcal{G}$}
        \IF{$X\not\in \tf{N}_Y$ and $Y\not\in \tf{N}_X$}
        \STATE{Delete the edge between $X$ and $Y$ in $\mathcal{G}$.}
        \ENDIF
        \ENDFOR
        \FOR{each edge $X -Y$}
            \IF{there is prior knowledge on the orientation $X - Y$}
            \STATE{Orient $X - Y$ according to the prior knowledge.}
            \ENDIF
        \ENDFOR
        \FOR{each pair of non-adjacent variables $(X, Y)$ with a common neighbor $Z$}
        \STATE{Find d-separation set $S(X,Y)$ with the largest $p$-value in the neighborhood of $X$ and $Y$.}
        \IF{$Z \not\in S(X,Y)$}
        \STATE{Orient $X - Z - Y$ as $X \rightarrow Z \leftarrow Y$.}
        \ENDIF
        \ENDFOR
        \STATE{Form $\mathcal{G}$ by recursive orientation according to the following two rules:\\
            1. If $X - Y$ and there is a directed path from $X$ to $Y$, then orient $X - Y$ as $X \rightarrow Y$;\\
            2. If $X$ and $Y$ are not adjacent and there is a $Z$ such that $X \rightarrow Z$ and $Z - Y$, then orient $Z - Y$ as $Z \rightarrow Y$.}
    \end{algorithmic}
\end{algorithm}
\end{singlespace}

\noindent\textcolor{black}{\tf{Remark 1}: Note that multiple orientations may satisfy the inferred d-separation (conditional independence) structure. Hence, in Algorithm~3, we first use prior knowledge in the orientation process to establish orientations and to enhance the interpretability of the orientation result. For instance, we presume that {age may lead to education status} but not the other way around.
The prior knowledge is provided in the sample code of the supplemental material.}

\noindent\textcolor{black}{\tf{Remark 2}: Note that in the d-separation set searching procedure in Algorithm~3, we use the d-separation set $S(X,Y)$ with the largest $p$-value to stabilize the orientation results. 
It agrees with equation~\eqref{eq:snxy} used in the maximization step for skeleton learning.
The approach of using the conditional independence set with the largest $p$-value in the orientation process has also been used in \citet{ramsey2016improving}.
Sections~2.5 and 2.6 both use the largest $p$-value among the conditional independence tests but for different purposes. Section~2.5 is about skeleton learning, and Section~2.6 is about orientation. \citet{ramsey2016improving} only concerned the orientation phase but not the skeleton learning.}

\noindent\tf{Remark 3}:
To make the graph learned by the proposed algorithm more interpretable, we calculate $p$-values for the significance of the (undirectional) connections for all edges in the graph learned by Algorithm~\ref{alg:structure_learn}.
More specifically, for edge $X-Y$ between the covariates $X$ and $Y$, define $P(X,Y)$ to measure the significance of $X - Y$ as follows
\begin{equation}
\label{eq:PXY}
P(X,Y) = \min\{S_{N_X}(X,Y), S_{N_Y}(X,Y)\},
\end{equation}
where $\tf{N}_X$ and $\tf{N}_Y$ are the neighbors of $X$ and $Y$, respectively.
Note that the measure $P(X,Y)$ is undirectional, i.e.,
the orientation of the edge $X - Y$ has no effect on $P(X,Y)$ from its definition,
and $P(X,Y)$ also has no information on the edge orientation.

\section{Application to MPHIA Data}
\label{section:Results}

The MPHIA survey is a new HIV-focused, cross-sectional, household-based, nationally representative survey of adults and adolescents aged 15 years and older as well as children aged 0-14 years. In addition to HIV testing results, the survey contains demographic questions, such as age group, gender (preference to the gender collected in the adult or adolescent questionnaire), ethnic group, and HIV-related questions, such as access to preventive care and treatment services. \textcolor{black}{There were 26,871 survey participants and 1,407 covariates in total. We will focus on adults and adolescents aged 15 years and older with complete Tri90 related information. There are 2,217 such individuals in the MPHIA survey data, including 712 males and 1,505 females. That is to say, these 2,217 individuals are HIV-positive with known status of HIV awareness (Aware), antiretroviral therapy (ART), and viral load suppression (VLS).}

\textcolor{black}{In PHIA surveys, the 1,407 covariates include 34 continuous variables such as age and time, 32 discrete variables such as the number of partners, and 1,341 categorical variables (nominal ones such as gender and ethnic group, ordinal ones such as alcohol frequency). Some covariates are applicable to males or females only, i.e., the questions related to pregnancy are only applicable to females. Therefore, the numbers of vertices are quite different between the female graph and the male graph. 
After a data-preprocessing procedure (illustrated later), we will apply the causal structural learning algorithm to six datasets for each combination of gender and 90-90-90 goal separately.}



\begin{singlespace}
\begin{table}
\caption{\label{tab:Table0} Sample sizes $n$ and numbers of covariates N(V) of the six datasets for each combination of gender and 90-90-90 goal after data preprocessing. Aware,  ART, and VLS stand for HIV awareness, ART treatment, and viral load suppression respectively. The Male/Female column means that the DAG is learned using only the male/female participants in MPHIA, and the rows represent the Tri90 goals.}
\centering
\begin{tabular}[t]{lrrrr}
\toprule
\multicolumn{1}{c}{ } & \multicolumn{2}{c}{Male} & \multicolumn{2}{c}{Female} \\
\cmidrule(l{3pt}r{3pt}){2-3} \cmidrule(l{3pt}r{3pt}){4-5}
Goals & $n$ & N(V) & $n$ & N(V)\\
\midrule
Aware & 712 & 66 & 1,505 & 93\\
ART & 510 & 66 & 1,210 & 92\\
VLS & 454 & 66 & 1,110 & 92\\
\bottomrule
\end{tabular}
\end{table}
\end{singlespace}


\noindent{\bf Overview of the data-preprocessing procedure}: 
We first drop all the covariates with a dominant level (one level has $>99\%$ samples), because the MPHIA sample size is not large enough to reject any null hypothesis of the conditional independence involving those variables. Also, we ``merge" all the closely related covariates into a single covariate in the MPHIA dataset. For example, some multi-option questions are dummy-coded by many two-option ones in the MPHIA dataset,
and we combine them together to create multi-level categorical covariates. By ``merging" these kinds of covariates, we can reduce the number of covariates and improve the interpretability of our results.
Also, some covariates appear in the MPHIA dataset multiple times with the same meaning but different names, such as EthnicGroup and EthnicCode, and we keep one covariate and drop the others in such situations.
We further remove some covariates that are direct indicators of the Tri90 goals from the MPHIA data. For example, there are questions like whether the subject takes a certain ART medicine or not in the MPHIA survey. These covariates are strongly correlated to the ART status, but not helpful for the purpose of building the causal pathways because they may block the connection between the ART status and other meaningful covariates.
Furthermore, some categorical variables have many levels which can complicate the analysis. To reduce the number of levels while keeping the main information of each covariate, we keep the biggest levels of each covariate which cover at least 95\% of the individuals, and combine the remaining levels into the ``Others" category.
\textcolor{black}{We sub-sample the whole dataset by gender and by Tri90 goal to create six datasets, and within each of the six datasets, we further drop the covariates with a dominant level (one level has $>99\%$ samples). The sample sizes and the numbers of vertices in each dataset are summarized in Table~\ref{tab:Table0}. There is one covariate included in the female awareness dataset but not in the female ART and VLS datasets: LiveHere (whether the individual lives here or not). 
The covariate is categorical and has a dominating level, which is below the threshold of 99\% in the female awareness dataset but is above the threshold in the female ART and VLS datasets. More details about the six datasets are provided in Supplement~\ref{section:6datasets}.}

In the next section, we compare the results obtained from the proposed algorithm with those of the existing algorithms in Section~\ref{section:Validation}. We further discuss the potential Tri90 pathways discovered by the proposed algorithm in detail in Section~\ref{section:Discussions}.


\subsection{Model Comparison}
\label{section:Validation}

We also apply the following existing structural learning algorithms to each dataset: 
\textbf{PC-stable} algorithm proposed by \citet{colombo2014order}, which is a stable/order-independent variant of the original PC (initials of the first names) algorithm proposed by \citet{spirtes1991algorithm};
\textbf{MMPC} (Max Min Parents and Children) algorithm proposed by \citet{tsamardinos2003mmpc}; 
\textbf{IAMB} (Incremental Association Markov Blanket) algorithm proposed by \citet{tsamardinos2003iamb}; 
\textbf{GS} (Grow-Shrink) algorithm proposed by \citet{margaritis2003gs}.
Notice that we use the PC-stable algorithm instead of the PC algorithm since the PC-stable algorithm is order-independent.
Order independence means that a random reordering of the variables does not affect the graphical learning result, which enhances the stability of the algorithm and also makes the results more interpretable.

\textcolor{black}{We use the existing graphical learning algorithms implemented in the \textbf{R} package \textbf{bnlearn} \citep{scutari2010bnlearn} in Sections~3 and 4.
The default conditional independence test statistic in the \textbf{bnlearn} package is the mutual information for categorical variables and the linear correlation for continuous variables.
We set the upper bound of sizes of conditional sets $M_{\text{CI}} = 3$ and the size of the conditional independence test $\alpha=0.05$ by default for all the causal structural learning algorithms used in the comparison.}

We first summarize the structural learning results by different graphical learning methods. 
\textcolor{black}{The results in Table~\ref{tab:Table1} shows that the proposed algorithm makes new discoveries regarding Tri90 goals.}
More specifically, Table~\ref{tab:Table1} show that the numbers of edges (NE) and the number of directed edges (NDE) are both much smaller than the number of vertices (NV) in the graphs learned by the existing algorithms including the PC-stable, the MMPC, the IAMB, and the GS algorithms. These existing algorithms lean toward fractured graphs and do not have much conditional independence information for the orientation. On the contrary, the proposed algorithm produces larger numbers of edges and well-connected graphs and has more potential to infer the directions of edges.

\begin{singlespace}

\begin{table}[!ht]
\caption{\label{tab:Table1}Structural learning results by different graphical learning methods for 90-90-90 goals. Aware, ART, and VLS stand for the three 90-90-90 targets of HIV awareness, ART treatment, and viral load suppression respectively.
NV, NE, and NDE stand for number of vertices, edges, and directed edges respectively.}
\centering
\begin{tabular}[t]{llrrrrrr}
\toprule
\multicolumn{2}{c}{ } & \multicolumn{3}{c}{Male} & \multicolumn{3}{c}{Female} \\
\cmidrule(l{3pt}r{3pt}){3-5} \cmidrule(l{3pt}r{3pt}){6-8}
Goals & Method & NV & NE & NDE & NV & NE & NDE\\
\midrule
 & PC-stable & 66 & 10 & 0 & 93 & 19 & 2\\

 & MMPC & 66 & 6 & 0 & 93 & 18 & 2\\

 & IAMB & 66 & 8 & 0 & 93 & 17 & 0\\

 & GS & 66 & 10 & 0 & 93 & 10 & 0\\

\multirow{-5}{*}{\raggedright\arraybackslash Aware} & New & 66 & 113 & 99 & 93 & 156 & 142\\
\cmidrule{1-8}
 & PC-stable & 66 & 3 & 0 & 92 & 19 & 2\\

 & MMPC & 66 & 3 & 0 & 92 & 15 & 0\\

 & IAMB & 66 & 4 & 0 & 92 & 17 & 0\\

 & GS & 66 & 11 & 0 & 92 & 12 & 0\\

\multirow{-5}{*}{\raggedright\arraybackslash ART} & New & 66 & 103 & 86 & 92 & 158 & 149\\
\cmidrule{1-8}
 & PC-stable & 66 & 1 & 0 & 92 & 20 & 4\\

 & MMPC & 66 & 4 & 0 & 92 & 15 & 0\\

 & IAMB & 66 & 4 & 0 & 92 & 15 & 0\\

 & GS & 66 & 10 & 0 & 92 & 13 & 0\\

\multirow{-5}{*}{\raggedright\arraybackslash VLS} & New & 66 & 104 & 90 & 92 & 153 & 146\\
\bottomrule
\end{tabular}
\end{table}

\end{singlespace}

\textcolor{black}{\noindent\tf{Remark}: It is important to note that the goal of the causal graphical algorithms is not to produce as many edges and directed edges as possible. Later in this section, we will use the Bayesian Information Criterion (BIC) to show that these discoveries made by the proposed algorithm provide useful information about the MPHIA data.
Furthermore, in Section~\ref{section:Discussions}, we discuss the discovered pathways in detail, which are reasonable, also confirmed in other HIV Tri90 literature, and can provide useful insight for the three Tri90 goals.}

Let $\mathbf{d}$ be the distance (defined as the length of shortest path regardless of direction) from a particular 90-90-90 goal (awareness of HIV, ART, or VLS) to a covariate and N($\mathbf{d} \leq k$) be the number of covariates whose distances to a 90-90-90 goal are smaller than or equal to $k$. Very few covariates are close to 90-90-90 goals in the graphs learned by the existing algorithms with only two covariates whose distance to the 90-90-90 goals are smaller or equal to three. On the contrary, the proposed algorithm discovers many covariates that are of a small distance to the 90-90-90 goals including several direct neighbors ($\mathbf{d} \leq 1$), which includes the ones discovered by the existing algorithms. See details in Supplement Table~\ref{tab:Table2}.




The Bayesian information criterion (BIC, \citep{BIC}) is a classical statistical tool for model selection.
We compare our proposed graphical learning algorithm with the aforementioned classical PC-stable, MMPC, IAMB, and GS algorithms using BIC criterion and summarize the results in Table~\ref{tab:IC}. Let $\tf{D}_k$, $k=1,2,\cdots,6$, be the data sets corresponding to the three 90-90-90 goals of each gender, respectively.
For $k=1,2,\cdots,6$,
we use each of the graphical learning algorithm $A_i$, for $i=1,2,\cdots,5$,
to learn a DAG $\mathcal{G}_{i,k}$ on $\tf{D}_{k}$,
where $A_i$, $i=1,2,\cdots,5$, stand for the PC-stable, MMPC, IAMB, GS, and the proposed algorithm, respectively.

The log-likelihood of a DAG $\mathcal{G}$ can be decomposed as follows:
\[
\ell(\theta|\mathcal{G})
=\frac{1}{n} \sum_{i=1}^n \log(p(x_{i1},\cdots,x_{ip}|\mathcal{G};\theta))
\]

\begin{equation}
= \sum_{j=1}^p \left[
    \frac{1}{n} \sum_{i=1}^n  \log(p(x_{ij}|\pi_{ij}(\mathcal{G});\theta_j)\right] 
= \sum_{j=1}^p \ell_j(\theta_j|\pi_{j}(\mathcal{G})),
\label{eqn:loglikelihood}
\end{equation}
where $\theta_j$'s are parameters of the model, $n$ is the sample size, $p$ is the number of covariates, $x_{ij}$, $1\leq i\leq n$, $1\leq j\leq p$, is the $i$-th observation of the $j$-th covariate $X_j$, $\pi_{j}(\mathcal{G})$ is the set of parents of $X_j$ in the DAG $\mathcal{G}$, and $\pi_{ij}(\mathcal{G})$ is the $i$-th observation of $\pi_{j}(\mathcal{G})$.
It shows that the log-likelihood of $\mathcal{G}$ can be decomposed as the sum of the log-likelihood of local structures of a covariate given its parents.
For the estimation of the local structures,
existing literature on causal graphical models often assumes a linear model of a covariate on its parents \citep{spirtes2010introduction, Valente2010searching, bolla2019graphical}.
Note that in the MPHIA data set, many covariates are categorical, so we assume generalized linear models (GLM) of a covariate given its parents instead of linear models, and we fit the GLM of local structures by MLE.
The degree of freedom (DF) and the log-likelihood of a DAG are the sums of DFs and log-likelihoods of the local structures, respectively.

Note that all the \textit{Log-likelihood*} in Table~\ref{tab:IC} are positive
since they are the differences between the log-likelihood and the log-likelihood of the null model (model with intercept only).
Furthermore, in Table~\ref{tab:IC},
BIC score is calculated by $-2\textit{Log-likelihood*} + \textit{DF} \log(n)$,
where $n$ is the sample size. Hence lower BIC scores correspond to better models.

\begin{singlespace}

\begin{table}[!ht]
\caption{\label{tab:IC}Comparison of different graphical learning methods by Bayesian Information Criterion. Aware, ART, and VLS stand for the three 90-90-90 targets of HIV awareness, ART treatment, and viral load suppression respectively. DF stands for the number of degree of freedom.
Log-likelihood* is the difference between the log-likelihood and the log-likelihood of the null model (model with intercept only).}
\centering
\resizebox{\linewidth}{!}{
\begin{tabular}[t]{llrrrrrr}
\toprule
\multicolumn{2}{c}{ } & \multicolumn{3}{c}{Male} & \multicolumn{3}{c}{Female} \\
\cmidrule(l{3pt}r{3pt}){3-5} \cmidrule(l{3pt}r{3pt}){6-8}
Goals & Method & DF & Log-likelihood* & BIC & DF & Log-likelihood* & BIC\\
\midrule
 & PC-stable & 50 & 3,076.7 & -5,825.0 & 142 & 12,051.1 & -23,063.3\\

 & MMPC & 44 & 2,218.7 & -4,148.5 & 140 & 12,853.6 & -24,683.0\\

 & IAMB & 71 & 2,650.4 & -4,834.5 & 134 & 14,391.3 & -27,802.2\\

 & GS & 81 & 2,348.3 & -4,164.7 & 69 & 6,027.7 & -11,550.6\\

\multirow{-5}{*}{\raggedright\arraybackslash Aware} & New & 948 & 17,755.5 & -29,284.5 & 1,012 & 52,807.1 & -98,209.8\\
\cmidrule{1-8}
 & PC-stable & 12 & 534.1 & -993.4 & 147 & 10,172.2 & -19,300.9\\

 & MMPC & 26 & 1,146.7 & -2,131.3 & 128 & 9,833.0 & -18,757.5\\

 & IAMB & 32 & 1,454.8 & -2,710.1 & 132 & 11,160.4 & -21,383.9\\

 & GS & 62 & 1,573.6 & -2,760.7 & 96 & 6,712.0 & -12,742.5\\

\multirow{-5}{*}{\raggedright\arraybackslash ART} & New & 970 & 10,274.1 & -14,500.7 & 2,117 & 48,597.2 & -82,167.2\\
\cmidrule{1-8}
 & PC-stable & 6 & 162.4 & -288.1 & 144 & 9,448.7 & -17,887.6\\

 & MMPC & 30 & 1,238.2 & -2,292.8 & 140 & 10,344.0 & -19,706.3\\

 & IAMB & 32 & 1,293.2 & -2,390.7 & 124 & 10,352.2 & -19,834.9\\

 & GS & 63 & 1,888.7 & -3,392.0 & 104 & 6,074.3 & -11,419.3\\

\multirow{-5}{*}{\raggedright\arraybackslash VLS} & New & 1,184 & 9,222.8 & -11,201.8 & 979 & 41,991.0 & -77,117.2\\
\bottomrule
\end{tabular}}
\end{table}

\end{singlespace}

Our proposed algorithm learns a much larger number of edges in all the six graphs compared with the existing algorithms in Table~\ref{tab:Table1}, and thus it has much larger degrees of freedom for the log-likelihood defined in Equation \eqref{eqn:loglikelihood}. Table~\ref{tab:IC} shows that the proposed algorithm has the largest degree of freedom 
and the largest log-likelihood. 
The much larger log-likelihoods imply that the proposed algorithm discovers a lot more useful information in MPHIA, which is further confirmed by the best (smallest) BIC scores in Table~\ref{tab:IC}.

\subsection{90-90-90 Pathways}
\label{section:Discussions}


Table~\ref{tab:Neighbors} lists the neighbors of the 90-90-90 goals discovered by the proposed causal structural learning algorithm.
Details of the graphs learned by the proposed algorithm are provided in Figures~\ref{fig:tri90awarefemale} and \ref{fig:tri90awaremale}|\ref{fig:tri90vlsmale}, which render parts of the graphs surrounding each 90-90-90 goal for each gender.
\textcolor{black}{The partial graphs we present in Figures~\ref{fig:tri90awarefemale} and \ref{fig:tri90awaremale}|\ref{fig:tri90vlsmale} are uniquely determined using our proposed algorithm with prior knowledge in the orientation process.
Hence the problem of multiple orientations does not affect the interpretation of the results of the MPHIA data. See Remark 1 of Algorithm~\ref{alg:structure_learn}.}
Also note that the covariates are renamed to have more intuitive meaning than their original names in MPHIA data codebooks. The MPHIA data codebook and the meaning of important covariates are provided in Supplement~\ref{section:Codebook}. Those covariates discovered by the proposed algorithm are consistent with findings in the existing literature, and we discuss them by Tri90 goals and by genders.
\textcolor{black}{
However, we should be cautious about the interpretation of the results of the proposed algorithm. Notably, the proposed algorithm assumes causal sufficiency, implying no hidden confounding (or common causes) covariates.
Although the MPHIA survey has included 1,407 related covariates, and most of the confounding covariates are likely observed, there may still be some unobserved confounding factors that can affect the interpretation of the proposed algorithm results.
}

\begin{singlespace}

\begin{table}[!ht]
\caption{\label{tab:Neighbors}Neighbors of 90-90-90 goals discovered by the proposed graphical learning method. Aware, ART, and VLS stand for the three 90-90-90 targets of HIV awareness, ART treatment, and viral load suppression respectively. Covariates in each table cell are arranged in alphabetical order. Code book can be found in Supplement~\ref{Codebook:neighbors}.}
\centering
\begin{tabular}[t]{lll}
\toprule
\multicolumn{1}{c}{ } & \multicolumn{1}{c}{Male} & \multicolumn{1}{c}{Female} \\
\cmidrule(l{3pt}r{3pt}){2-2} \cmidrule(l{3pt}r{3pt}){3-3}
Goals & Neighbors of goal & Neighbors of goal\\
\midrule
Aware & \makecell[l]{AlcoholFrequency,\\PartnerAge,\\PartnerNumber12Mo,\\PLWHSupportGroup,\\ViolenceOK?,\\WifeNumLiveElsewhere} & \makecell[l]{AgeGroup,\\EasyGetCondom,\\Education,\\PLWHSupportGroup,\\PregNum}\\
\cmidrule{1-3}
ART & \makecell[l]{AbnormPenisDischarge,\\PartnerNumber12Mo,\\TravelTime,\\WifeNumLiveElsewhere} & \makecell[l]{SyphilisTestInPreg,\\TravelTime,\\ViolenceOK?,\\WifeNumOfHusband}\\
\cmidrule{1-3}
VLS & \makecell[l]{SeekMedicalHelp,\\WifeNum} & \makecell[l]{ForceSexTimes,\\SupportGroupTimes12Mo,\\TranslatorUsed}\\
\bottomrule
\end{tabular}
\end{table}

\end{singlespace}

\subsubsection{HIV awareness among females}
\begin{figure}[!ht]
    \begin{center}
    \begin{adjustbox}{center}
        \includegraphics[width=10in,height=6.6in]{"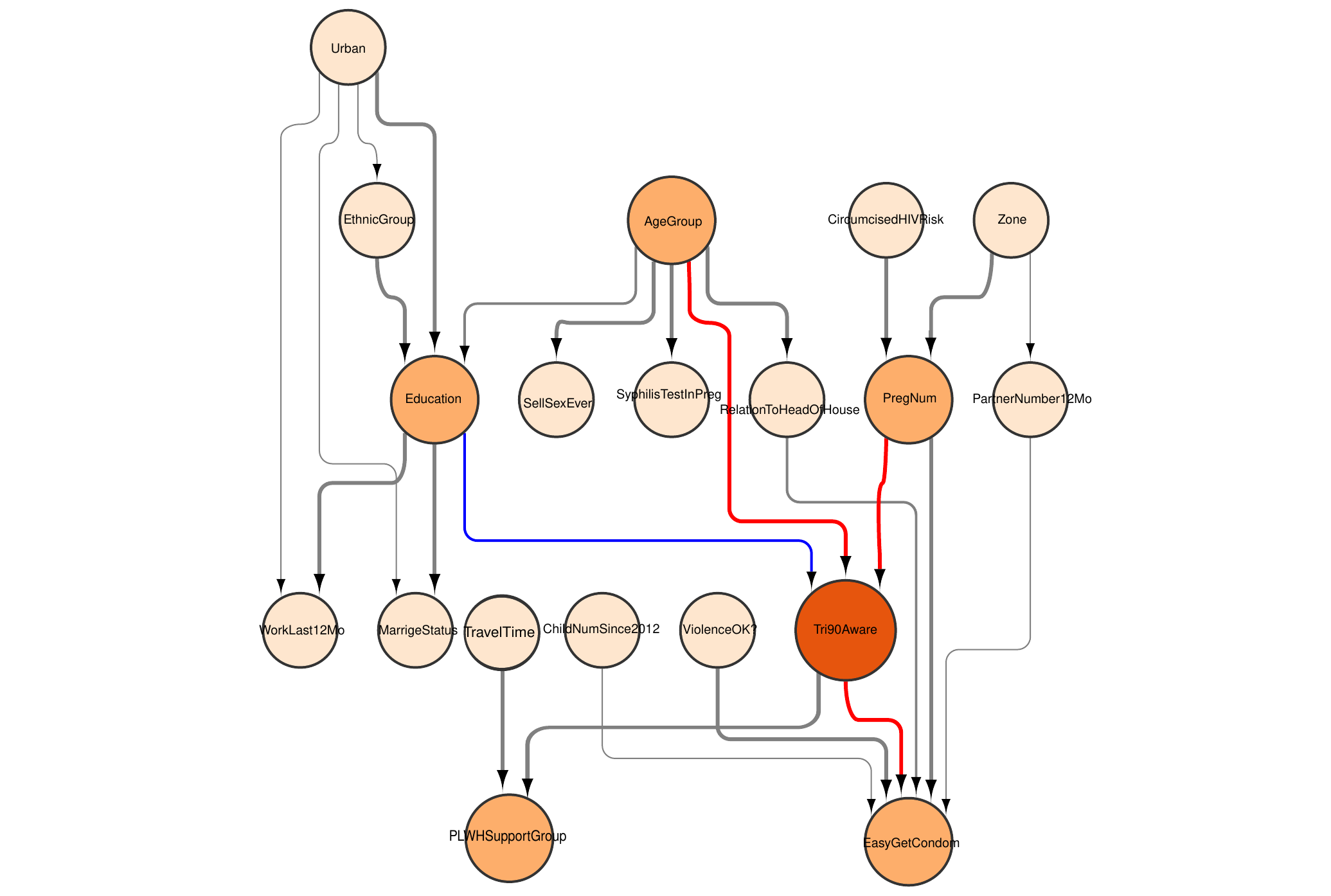"}

        \end{adjustbox}
        \begin{singlespace}
        \caption{90-90-90 Awareness graph in female.
        Vertices representing the Tri90 goals are biggest and marked by orange; vertices closer to goals have bigger sizes and darker colors than those farther away from goals. Widths of edges reflect the significance of the non-directional connection (conditional dependence) between vertices.
        Red and blue edges represent positive and negative relationships with Tri90 goals, respectively.
        Codebook can be found in Supplement~\ref{Codebook:tri90awarefemale}.}
        \label{fig:tri90awarefemale}
        \end{singlespace}
    \end{center}
\end{figure}

For HIV awareness among females, we use Figure~\ref{fig:tri90awarefemale} to visually illustrate the covariates that are closely connected to HIV awareness (distances $\leq 2$).
The learned graphs for other Tri90 goals and genders are provided in Supplement~\ref{section:TableFigure} Figures~\ref{fig:tri90awaremale}|\ref{fig:tri90vlsmale}.


We find that HIV-positive females with more pregnant times are more likely to know their HIV status. We think this is because pregnant women are more likely to be tested for HIV during antenatal clinic visits or laboring. \citet{Peltzer2009} reported that age, condom usage, and education are associated with HIV status awareness. The proposed algorithm finds that AgeGroup and Education are potential ``reasons" for female's HIV awareness, and that EasyCondom and PLWHSupportGroup may be results of female's HIV awareness: (1) older individuals are more likely to know their HIV status; (2) individuals who are unaware of their HIV status often do not know whether it is easy to get condoms or have difficulties in getting condoms; (3) females unaware of their status never answer the PLWHSupportGroup question; (4) individuals with higher education are less likely to be aware of their HIV positive status, which is opposite of the marginal association reported in \citet{Peltzer2009}. It is because (4) is the conditional association given AgeGroup (which affects both Education and HIV awareness): while females with higher education levels in the age group 35 to 44 are more likely to be aware of HIV positive status, females with higher education levels in other age groups are less likely to be aware of the HIV status. Further studies will be needed to establish the relationship between education and HIV status awareness.



Figure~\ref{fig:tri90awarefemale} also presents other covariates that indirectly connect to the HIV awareness among female HIV patients. Among those, employment status (WorkLast12Mo) and marital status are impacted by both urban residence and education; and education is impacted by ethnic group, urban residence, and age group.
AgeGroup is a potential ``reason" for SellSexEver and SyphilisTestInPreg: 
younger generations are less likely to have SyphilisTestInPreg; older generations are less likely to have sold sex ever.
CircumcisedHIVRisk is a score measuring whether the individual agrees that men who are circumcised are not at risk of HIV at all, do not need to use condoms, and can have multiple sexual partners without the risk of HIV. The graph suggests 
females that are more cautious toward HIV risk (lower CircumcisedHIVRisk score) have smaller numbers of pregnancies on average. 
Finally, more sexual partners (PartNumber12Mo), being a partner to the head of house (RelationToHeadOfHouse), and lower tolerance towards violence (ViolenceOK) lead to easiness in getting condoms.


\subsubsection{HIV awareness among males}
We find males whose partners are older are more likely to be aware of their HIV status, both marginally, or conditioning on their age groups. Comparing the male awareness pathway and the female awareness pathway, we suspect that older females contribute more to HIV awareness because someone's age group is associated with his/her partner's age. 

In Figure~\ref{fig:tri90awaremale}, the proposed algorithm finds that PartnerNumber12Mo (sexual partners in the last 12 months) and WifeNumLiveElsewhere (number of wive/partners that live elsewhere) to be important covariates for male HIV awareness. These discoveries are supported by \citet{Peltzer2009} which showed that HIV awareness is significantly associated with the number of sexual partners. 
WifeNumLiveElsewhere and PartnerNum12Mo are strongly correlated and have closely related meanings.
Instead of studying their impacts on HIV awareness separately,
we study the relationship between HIV awareness and WealthQuintile which is the parent of both WifeNumLiveElsewhere and PartnerNum12Mo.
We find that ``poor" HIV positive males (with wealth in lower 20\%) are
more likely to have no partner in the last 12 months, more likely to have wives/partners live elsewhere, and less likely to be aware of their HIV status than the wealthier (in the upper 80\%) HIV positive males. This finding is consistent with \citet{dokubo2014awareness}.

We also find that ViolenceOK and AlcoholFrequency may be possible ``reasons" of HIV awareness in males, where ViolenceOK is a score to measure whether the individual believes it is right for a man to beat his wife/partner under various scenarios.
Further investigation shows that among males who are HIV positive: (1) those with a higher violence score are more likely to be unaware of their HIV status; (2) those who never drink are most likely to be aware of their HIV status while those with a high frequency of alcohol use (more than four drinks a week) are least likely to be aware of their HIV positive status, and such an effect is more significant in males who have not worked in the last 12 months. Alcohol usage is an important factor associated with sexual risk behavior \citep{Kalichman2007, Hahn2011}, and violence score can also be an important factor for sexual behavior. These factors deserve more attention when advocating HIV awareness.

The proposed causal structural learning algorithm also finds that WorkLast12Mo has a small distance to HIV awareness which is supported by \citet{Peltzer2009}. PLWHSupportGroup is a result of HIV awareness because males unaware of their status never answer the PLWHSupportGroup question.


\subsubsection{ART pathways}
Next, we discuss the ART pathways for females and males in one subsection because they are relatively simpler compared with the HIV awareness pathways. For the same reason, we discuss the VLS pathways for females and males in the same subsection.

\citet{Hodgson2014}, \citet{Tomori2014}, and \citet{Kebaabetswe2019} showed the connection between TravelTime to the receiving of ART. The proposed algorithm indicates that among people living with HIV and being aware of their positive HIV status, those who do not know the travel time to HIV care are less likely to receive ART treatment than the individuals who know the travel time.

\citet{Friedman2015, Hatcher2015, Sullivan2015} showed that violence and other kinds of abuse are significantly associated with ART initiation. We find females with smaller violence scores are more likely to be on ART treatment.

We find that wealthier males are more likely to initiate the ART treatment, which is in agreement with \citet{gebru2018perceived}. On the other hand, the males with ``poor" wealth are likely to have no partners in the last 12 months, and more likely to have wives or partners live elsewhere.
The proposed algorithm also discovers that females with known status of SyphilisTestInPreg are more likely to be on ART treatment. \textcolor{black}{The learned female ART pathway reveals that those  received ART treatments had easy access to the HIV care (short travel time) and easy access to antenatal care where they could be offered the syphilis test during pregnancy. They were both indicators of receiving good health care services which increased the chance of receiving ART treatment.}


Finally, we find AbnormPenisDischarge to be a potential ``reason" for male ART initiation. Further investigation shows that males with the missing response on abnormal penis discharge problems are more likely to be on ART treatment among the HIV-positive males who are aware of their HIV status. Unfortunately, it was unclear why people did not respond AbnormPenisDischarge question.



\subsubsection{VLS pathways}

We find females who attend the support groups more frequently are more likely to have their viral load suppressed. It is supported by \citet{Roberts2000} \citet{Tomori2014}, and \citet{Rangarajan2016}.


\citet{Friedman2015, Hatcher2015, Sullivan2015} showed that violence and other kinds of abuse are significantly associated with viral load suppression (VLS), especially among females. It supports the connection between VLS and ForceSexTimes found in the female VLS pathway. 


We also find SeekMedicalHelp and WifeNum to be potential ``reasons" of VLS among males who are on ART. Males who seek help from doctors or nurses because of health issues such as abnormal penis discharge and painful urination are more likely to have viral load unsuppressed; and males with more wives are more likely to have viral load suppressed.

\section{Simulation Studies}
\label{section:Simulations}

In Section~3, we see that the proposed causal structural learning algorithm discovers many more new edges than the existing ones, and we validate the results of the proposed algorithm through the Bayesian information criterion (BIC).
However, since the true causal graph for MPHIA data is unknown, we cannot validate the edge discoveries directly.
To get more insights on the Type I and Type II error rates and their trade-off for the proposed algorithm against existing ones, we carry out simulation studies with settings mimicking the MPHIA data in this section.

\textcolor{black}{We check the true positive and negative rates of the proposed algorithm against the existing algorithms on synthetic data sets that mimic the MPHIA data. More specifically, we use the DAG learned from MPHIA data as the truth to generate the simulation data.}
Here we choose the graphs learned by the proposed algorithm as the truth for simulation purposes since
Table~\ref{tab:IC} shows that the graphs learned by the proposed algorithm are better fits for the MPHIA data in the BIC criterion than those learned by the other algorithms.
That is to say, let $\mathcal{G}_k$ be the DAG learned by the proposed algorithm on the 90-90-90 MPHIA data set $\tf{D}_k$ for $k=1,2,\cdots,6$.
Then we fit the data distribution $\mathcal{P}_k$ based on $\mathcal{G}_k$ on the data $\tf{D}_k$.
We further randomly generate simulated data sets based on the distribution $\mathcal{P}_k$ with sample size $n$.
\textcolor{black}{The distribution estimation and the random data sets generation utilize the {\it bn.fit} (Bayesian network fitting) and {\it rbn} (random Bayesian network) functions in the \textbf{R} package \textbf{bnlearn} \citep{scutari2010bnlearn}.}
We then carry out the proposed algorithm together with the aforementioned PC-stable, GS, MMPC, and IAMB algorithms on the generated data sets.
Finally, we calculate true positive rates and true negative rates of edges disregarding the orientation for each algorithm.

\begin{singlespace}

\begin{table}

\caption{\label{tab:newsimulation}Empirical true positive rates and true negative rates of different causal structural learning algorithms (in percentage). Aware, ART, and VLS stand for the 90-90-90 targets of HIV awareness, ART treatment, and viral load suppression respectively.}
\centering
\resizebox{\linewidth}{!}{
\begin{tabular}[t]{llrrrrrrrrrrr}
\toprule
\multicolumn{3}{c}{ } & \multicolumn{5}{c}{True Positive Rate} & \multicolumn{5}{c}{True Negative Rate} \\
\cmidrule(l{3pt}r{3pt}){4-8} \cmidrule(l{3pt}r{3pt}){9-13}
Goals & Gender & $n$ & PC-stable & MMPC & IAMB & GS & New & PC-stable & MMPC & IAMB & GS & New\\
\midrule
 &  & 250 & 10.6 & 12.6 & 12.7 & 10.6 & 36.7 & 100.0 & 99.9 & 99.8 & 99.9 & 98.0\\

 &  & 500 & 14.6 & 16.0 & 15.2 & 12.0 & 43.0 & 100.0 & 99.9 & 99.9 & 99.9 & 98.0\\

 & \multirow{-3}{*}{\raggedright\arraybackslash Male} & 1,000 & 18.8 & 20.7 & 15.4 & 14.5 & 48.9 & 99.9 & 99.9 & 99.8 & 99.9 & 98.0\\

 &  & 250 & 14.7 & 13.9 & 14.4 & 10.0 & 32.9 & 100.0 & 100.0 & 99.9 & 99.9 & 98.4\\

 &  & 500 & 18.4 & 17.7 & 17.7 & 12.3 & 38.2 & 100.0 & 100.0 & 100.0 & 99.9 & 98.4\\

\multirow{-6}{*}{\raggedright\arraybackslash Aware} & \multirow{-3}{*}{\raggedright\arraybackslash Female} & 1,000 & 24.2 & 22.5 & 22.1 & 14.4 & 45.4 & 100.0 & 100.0 & 100.0 & 99.9 & 98.3\\
\cmidrule{1-13}
 &  & 250 & 10.6 & 13.3 & 12.3 & 11.0 & 36.9 & 100.0 & 99.9 & 99.9 & 99.9 & 98.0\\

 &  & 500 & 15.2 & 17.2 & 15.6 & 14.2 & 46.1 & 100.0 & 99.9 & 99.9 & 99.9 & 97.9\\

 & \multirow{-3}{*}{\raggedright\arraybackslash Male} & 1,000 & 21.0 & 22.9 & 20.7 & 17.5 & 51.6 & 100.0 & 99.9 & 99.9 & 100.0 & 97.7\\

 &  & 250 & 14.1 & 11.3 & 10.3 & 11.2 & 31.7 & 100.0 & 100.0 & 99.9 & 99.9 & 98.2\\

 &  & 500 & 18.4 & 12.6 & 9.4 & 13.4 & 37.5 & 100.0 & 100.0 & 99.9 & 99.9 & 98.1\\

\multirow{-6}{*}{\raggedright\arraybackslash ART} & \multirow{-3}{*}{\raggedright\arraybackslash Female} & 1,000 & 21.2 & 17.4 & 8.4 & 14.9 & 47.4 & 100.0 & 100.0 & 100.0 & 100.0 & 98.1\\
\cmidrule{1-13}
 &  & 250 & 9.4 & 13.1 & 13.0 & 11.0 & 35.7 & 99.9 & 99.9 & 99.9 & 100.0 & 97.9\\

 &  & 500 & 12.6 & 17.3 & 17.3 & 12.9 & 45.0 & 99.9 & 99.9 & 99.9 & 100.0 & 97.9\\

 & \multirow{-3}{*}{\raggedright\arraybackslash Male} & 1,000 & 16.3 & 21.5 & 21.3 & 16.3 & 49.6 & 99.9 & 99.9 & 99.9 & 100.0 & 97.5\\

 &  & 250 & 15.5 & 12.8 & 11.3 & 12.5 & 34.1 & 100.0 & 100.0 & 99.9 & 100.0 & 98.4\\

 &  & 500 & 19.6 & 13.8 & 10.0 & 14.3 & 39.9 & 100.0 & 100.0 & 99.9 & 100.0 & 98.4\\

\multirow{-6}{*}{\raggedright\arraybackslash VLS} & \multirow{-3}{*}{\raggedright\arraybackslash Female} & 1,000 & 23.5 & 17.6 & 10.1 & 15.0 & 46.0 & 100.0 & 100.0 & 99.9 & 99.9 & 98.3\\
\bottomrule
\end{tabular}}
\end{table}

\end{singlespace}

Here we set the sample size $n = 250, 500, 1000$ for a sample size similar with our real data and to check the performance of the proposed algorithm with different sample sizes.
We repeat the Monte Carlo simulation 500 times for each setting and summarize the results in Table~\ref{tab:newsimulation}.
The left and right panels of Table~\ref{tab:newsimulation} summarize the empirical true positive and negative rates of the proposed algorithm as well as those of existing algorithms, respectively.
From the right panel of Table~\ref{tab:newsimulation}, we can see that the proposed algorithm has similar true negative rates with existing algorithms.
Furthermore, from the left panel of Table~\ref{tab:newsimulation}, we can see that the proposed algorithm has better true positive rates than existing algorithms.

\textcolor{black}{We also present additional simulation studies in Supplement~\ref{section:AdditionalSimulation} to save space.}

\section{Conclusions}
\label{section:Conclusions}



UNAIDS 90-90-90 goals are important milestones to end AIDS.
To understand the progress on the 90-90-90 goals better, we analyze the Malawi PHIA (MPHIA) data set to discover important covariates and potential causal pathways for the 90-90-90 goals through causal structural learning in the paper.
Existing classical constrained-based causal structural learning algorithms are quite aggressive in edge removal and can lead to information losses while building directed graphical models, \textcolor{black}{especially in the case of categorical variables and relatively small sample sizes}.
To deal with the problem, we propose a new causal structural learning algorithm.
The proposed algorithm can preserve more information about important features and potential causal pathways as shown by various numerical studies \textcolor{black}{when many covariates in the domain are  categorical}.
More specifically, the proposed algorithm improves true positive rates over the existing classical algorithms while having a comparable true negative rate. It shows that our proposed algorithm has a great potential to discover important features and potential causal pathways, \textcolor{black}{especially in a domain with many categorical variables}.

Carrying out the causal graphical analysis on the MPHIA data set, we obtain interesting results on important covariates and possible causal pathways related to the UNAIDS 90-90-90 goals.
For example, the proposed algorithm discovers age and  condom usage to be important covariates for female HIV awareness and number of sexual partners to be important for male HIV awareness, which agrees with literature, such as \citet{Peltzer2009}.
The proposed algorithm also discovers similarities as well as differences between female and male pathways.
For example, travel time is discovered to be an important covariate for both female and male ART.
However, there are also different important covariates for female and male ART, such as attitude towards violence for female ART and partner numbers for male ART.

\textcolor{black}{It is also important to pay attention to the assumptions of the proposed causal structural learning algorithm when we interpret the results.
One important assumption behind the proposed algorithm is causal sufficiency which is critical for our algorithm and many other constrained-based causal structural learning algorithms. Although the MPHIA survey provides many related covariates, the causal sufficiency assumption may still not hold perfectly. }

\textcolor{black}{In the paper, we stratify the MPHIA dataset by sex and learn the graphical models for each sex and each Tri90 goal.
It is also possible to learn the DAG with further stratified data by other covariates such as age, but the sample size would be too small for each stratum to make reliable inferences. }


The discoveries on causality are important extensions for existing literature where only correlation instead of causation is established.
More studies can be carried out to further validate the potential causal discoveries, and other statistical inference tools such as mediation analysis can be applied to further the understanding of the causal relationship.
The discoveries on causality can help develop better HIV response strategies and related policies.


\begin{singlespace}
\bibliography{recursive}
\end{singlespace}

\newpage

\renewcommand{\baselinestretch}{1.2}

\setcounter{page}{1} 

\setcounter{equation}{0}
\setcounter{section}{0}
\setcounter{subsection}{0}
\setcounter{table}{0}
\setcounter{figure}{0}
\setcounter{theorem}{0}

\renewcommand{\thepage}{S.\arabic{page}}
\renewcommand{\theequation}{S.\arabic{equation}}
\renewcommand{\thesection}{S.\arabic{section}}
\renewcommand{\thesubsection}{S.\arabic{section}.\arabic{subsection}}
\renewcommand{\thetable}{S.\arabic{table}}
\renewcommand{\thefigure}{S.\arabic{figure}}
\renewcommand{\thetheorem}{S.\arabic{theorem}}

\centerline{SUPPLEMENT TO ``Causal Structural Learning on MPHIAIndividual Dataset"}

\

\



We first present proofs of Propositions~\ref{prop:cxs} and \ref{prop:snxy}, and then present
some additional details for the six Tri90 datasets, also some additional numerical results of Section 3 and parts of MPHIA Codebook.

\section{Proof of Proposition~\ref{prop:cxs}}
\label{section:proof_cxs}

It is easy to get \eqref{eq:initialC}, \eqref{eq:emptyC}, and \eqref{eq:inclusion} directly from the definition of $\tf{C}_X(\tf{S})$.
Furthermore, notice that \eqref{eq:upperbound} can be easily derived from \eqref{eq:inclusion}.
Hence the only thing that still needs to be proved is \eqref{eq:incremental_test}.

Suppose $\tf{S} = \{S_1, \cdots, S_n\}$, $n \geq 1$,
and $\tf{S}$ satisfies \eqref{eq:1}.
For any $C \in \tf{C}_X(\tf{S})$, we have $C \in \tf{C}^{*}_X(\tf{S}) \cap \tf{L}_X(\tf{S})$,
and we also have
$C\not\perp X|\tf{S}$, and $S_i\not\perp X| (\tf{S}_{-i} \cup \{C\})$, $i=1,\cdots,n$, by definition of $\tf{C}_X(\tf{S})$.
Hence we have $\tf{C}_X(\tf{S}) \subseteq
\{C \in \tf{C}_X^{*}(\tf{S}) \cap \tf{L}_X(\tf{S}),
C\not\perp X|\tf{S},
S_i\not\perp X| (\tf{S}_{-i} \cup \{C\}), i=1,\cdots,n
\}$.

Furthermore, we want to prove that $\tf{C}_X(\tf{S}) =
\{C \in \tf{C}_X^{*}(\tf{S}) \cap \tf{L}_X(\tf{S}),
C\not\perp X|\tf{S},
S_i\not\perp X| (\tf{S}_{-i} \cup \{C\}), i=1,\cdots,n
\}$ by contradiction.
If $\tf{C}_X(\tf{S}) \subsetneq
\{C \in \tf{C}_X^{*}(\tf{S}) \cap \tf{L}_X(\tf{S}),
C\not\perp X|\tf{S},
S_i\not\perp X| (\tf{S}_{-i} \cup \{C\}), i=1,\cdots,n
\}$, then there exists $C_0 \in \tf{C}_X^{*}(\tf{S}) \cap \tf{L}_X(\tf{S})$ such that $C_0\not\perp X|\tf{S}$, $S_i\not\perp X| (\tf{S}_{-i} \cup \{C_0\})$, $i=1,\cdots,n$, and $C_0 \notin \tf{C}_X(\tf{S})$.

From the definition of $\tf{C}_X(\tf{S})$ and $C_0 \in \tf{L}_X(\tf{S})$, the only way for $C_0$ not to be in $\tf{C}_X(\tf{S})$ is for $\{C_0\} \cup \tf{S}$ to violate \eqref{eq:1}.
So there exists $N_0\in (\{C_0\} \cup \tf{S})$ and $\tf{S}_0 \subseteq ((\{C_0\} \cup \tf{S}) \backslash \{N_0\})$ such that $N_0 \perp X | \tf{S}_0$.
\begin{enumerate}
    \item
    If $\tf{S}_0 \subsetneq ((\{C_0\} \cup \tf{S}) \backslash \{N_0\})$,
    then $(\{N_0\}\cup\tf{S}_0) \subsetneq (\{C_0\} \cup \tf{S})$.
    So there must exists $i_0$, $1\leq i_0 \leq n$, such that $(\{N_0\}\cup\tf{S}_0) \subseteq (\{C_0\} \cup \tf{S}_{-i_0})$,
    or $(\{N_0\}\cup\tf{S}_0) \subseteq \tf{S}$.
    Note that from the construction of $N_0$ and $\tf{S}_0$,
    we know $(\{N_0\}\cup\tf{S}_0)$ does not satisfy \eqref{eq:1}.
    Furthermore, any set with $(\{N_0\}\cup\tf{S}_0)$ as a subset does not satisfy \eqref{eq:1}.
    So $\{C_0\} \cup \tf{S}_{-i_0}$ or $\tf{S}$ does not satisfy \eqref{eq:1}, which is in contradiction with $C_0 \in \tf{C}_X(\tf{S}_{-i_0})$ and $\tf{S}$ satisfies \eqref{eq:1}.

    \item
    Hence we have $\tf{S}_0 = ((\{C_0\} \cup \tf{S}) \backslash \{N_0\})$.

    \begin{enumerate}
        \item
        If $N_0 = C_0$, then $\tf{S}_0=\tf{S}$ and $C_0 \perp X|\tf{S}$, which is in contradiction with $C_0\not\perp X|\tf{S}$.
        \item
        If $N_0 \neq C_0$, then there exists $i_0$, $1\leq i_0\leq n$, such that $N_0=S_{i_0}$.
        Then we have $\tf{S}_0 = \{C_0\}\cup\tf{S}_{-i_0}$,
        and $S_{i_0} \perp X|(\{C_0\}\cup\tf{S}_{-i_0})$, which is in contradiction with $S_{i_0} \not\perp X|(\{C_0\}\cup\tf{S}_{-i_0})$.
    \end{enumerate}
    In sum, we finish the proof of \eqref{eq:incremental_test} and Proposition~\ref{prop:cxs}.

\end{enumerate}

\section{Proof of Proposition~\ref{prop:snxy}}
\label{section:proof_snxy}

It is easy to get \eqref{eq:initialS} and \eqref{eq:inclusionS} directly from the definition of $S_N(X,Y)$.
Furthermore, notice that \eqref{eq:lowerboundS} can be easily derived from \eqref{eq:inclusionS}.
Hence the only thing that still needs to be proved is \eqref{eq:incremental_calcS}.

From the definition of $S_N(X,Y)$, we know that
$S_N(X,Y) \geq \CI(X,Y|\tf{N})$.
Hence we have
$S_N(X,Y) \geq \max\{S^{*}_N(X,Y), \CI(X,Y|\tf{N})\}$.
Suppose $S_N(X,Y) = \CI(X,Y,\tf{N}_0)$, where $\tf{N}_0 \subseteq \tf{N}$.
\begin{enumerate}
    \item
    If $\tf{N}_0 = \tf{N}$, then
    $S_N(X,Y) = \CI(X,Y|\tf{N}) \leq \max\{S^{*}_N(X,Y),\CI(X,Y|\tf{N})\}$.

    \item
    If $\tf{N}_0 \subsetneq \tf{N}$,
    then from the construction of $S^{*}_N(X,Y)$, we know that $S_N(X,Y) \leq S^{*}_N(X,Y) \leq \max\{S^{*}_N(X,Y), \CI(X,Y|\tf{N})\}$.

\end{enumerate}
In sum, we have
$$S_N(X,Y) \leq \max\{S^{*}_N(X,Y), \CI(X,Y|\tf{N})\}.$$
Furthermore, from $S_N(X,Y) \geq \max\{S^{*}_N(X,Y), \CI(X,Y|\tf{N})\}$, we have
$$S_N(X,Y) = \max\{S^{*}_N(X,Y), \CI(X,Y|\tf{N})\}.$$
Hence we finish the proof of Proposition~\ref{prop:snxy}.

\section{Additional Details of Six Tri90 Datasets by Target and Gender}
\label{section:6datasets}

\begin{enumerate}
    \item Aware: Among the 2,217 individuals included in our analysis, there are 1,720 individuals with self-reported awareness or antiretroviral (ARV) detected including 510 males and 1,210 females. So $\frac{1720}{2217} = 77.6\%$ MPHIA participants have achieved the first Tri90 goal \,---\, being aware of HIV status. We investigate important covariates and potential causal pathways for HIV awareness for males and females, respectively.
    \item ART: Among the 1,720 individuals with self-reported awareness or ARV detected, there are 1,564 individuals with self-reported ART or ARV detected including 454 males and 1,110 females. So $\frac{1564}{1720} = 90.3\%$ individuals have met the second Tri90 goal \,---\, being treated. We investigate important covariates and potential causal pathways for ART coverage for males and females, respectively.
    \item VLS: Among the 1,564 individuals with self-reported ART or ARV detected, there are 1,428 individuals with viral load suppression (VLS) including 408 males and 1,020 females. So $\frac{1428}{1564} = 91.3\%$ individuals have met the third Tri90 goal \,---\, reaching Viral Suppression. We investigate important covariates and potential causal pathways for VLS in males and females, respectively.
\end{enumerate}

\section{Additional Tables and Figures}
\label{section:TableFigure}


\begin{table}[!ht]

\caption{\label{tab:Table2}Number of important covariates for 90-90-90 goals discovered by different graphical learning methods. Aware, ART, and VLS stand for the three 90-90-90 targets of HIV awareness, ART treatment, and viral load suppression respectively.
$\mathbf{d}$ is the distance from  a  particular 90-90-90 goal (awareness of HIV, ART, or VLS) to a covariate, and $\text{N}(\mathbf{d}\leq k)$, $k=1,2,3$, are the number of covariates whose distances to a 90-90-90 goal are smaller than or equal to $k$.}
\centering
\resizebox{\linewidth}{!}{
\begin{tabular}[t]{llrrrrrr}
\toprule
\multicolumn{2}{c}{ } & \multicolumn{3}{c}{Male} & \multicolumn{3}{c}{Female} \\
\cmidrule(l{3pt}r{3pt}){3-5} \cmidrule(l{3pt}r{3pt}){6-8}
Goals & Method & N($\mathbf{d} \leq 1$) & N($\mathbf{d} \leq 2$) & N($\mathbf{d} \leq 3$) & N($\mathbf{d} \leq 1$) & N($\mathbf{d} \leq 2$) & N($\mathbf{d} \leq 3$)\\
\midrule
 & PC-stable & 0 & 0 & 0 & 0 & 0 & 0\\

 & MMPC & 1 & 1 & 1 & 0 & 0 & 0\\

 & IAMB & 0 & 0 & 0 & 0 & 0 & 0\\

 & GS & 0 & 0 & 0 & 0 & 0 & 0\\

\multirow{-5}{*}{\raggedright\arraybackslash Aware} & New & 6 & 19 & 49 & 5 & 18 & 51\\
\cmidrule{1-8}
 & PC-stable & 0 & 0 & 0 & 0 & 0 & 0\\

 & MMPC & 0 & 0 & 0 & 1 & 1 & 1\\

 & IAMB & 0 & 0 & 0 & 0 & 0 & 0\\

 & GS & 0 & 0 & 0 & 0 & 0 & 0\\

\multirow{-5}{*}{\raggedright\arraybackslash ART} & New & 4 & 13 & 38 & 4 & 11 & 28\\
\cmidrule{1-8}
 & PC-stable & 0 & 0 & 0 & 0 & 0 & 0\\

 & MMPC & 0 & 0 & 0 & 0 & 0 & 0\\

 & IAMB & 0 & 0 & 0 & 0 & 0 & 0\\

 & GS & 0 & 0 & 0 & 0 & 0 & 0\\

\multirow{-5}{*}{\raggedright\arraybackslash VLS} & New & 2 & 7 & 19 & 3 & 8 & 28\\
\bottomrule
\end{tabular}}
\end{table}

\clearpage

\KOMAoptions{paper=a3}
\recalctypearea
\newgeometry{top=20mm, left = 4in, bottom=-3in}

\begin{figure}[!ht]
    \begin{center}
    \begin{adjustbox}{center}
        \includegraphics[width=8.1in,height=13in]{"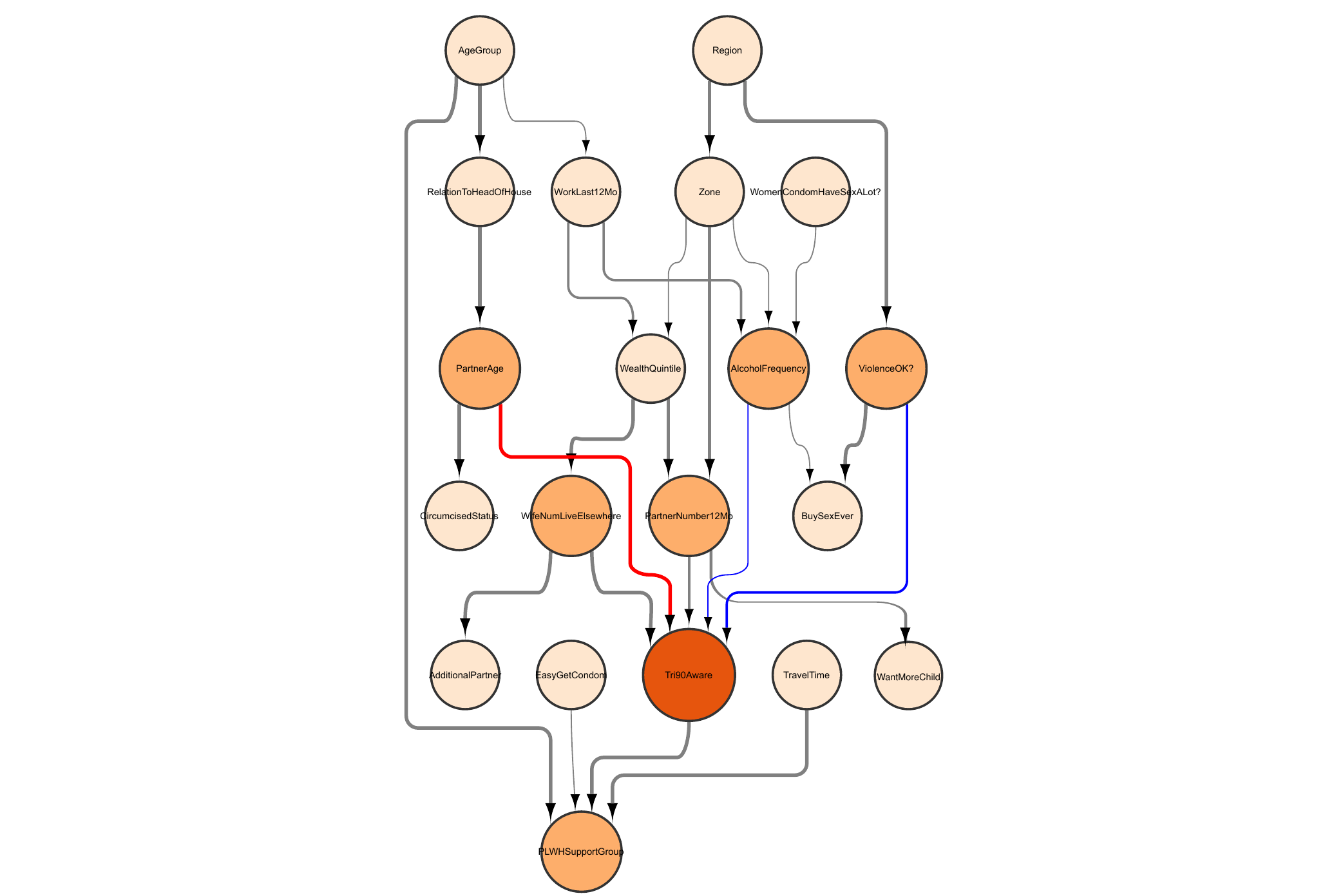"}
        \end{adjustbox}
        \captionsetup{width=9in}
        \caption{90-90-90 Awareness graph in male.
        Vertices representing the Tri90 goals are biggest and marked by orange; vertices closer to goals have bigger sizes and darker colors than those farther away from goals. Widths of edges reflect the significance of the non-directional connection (conditional dependence) between vertices.
        Red and blue edges represent positive and negative relationships with Tri90 goals, respectively.
        Codebook can be found in Supplement~\ref{Codebook:tri90awaremale}.}
        \label{fig:tri90awaremale}
    \end{center}
\end{figure}

\begin{figure}[!ht]
    \begin{center}
        \begin{adjustbox}{center}
        \includegraphics[width=13in,height=10in]{"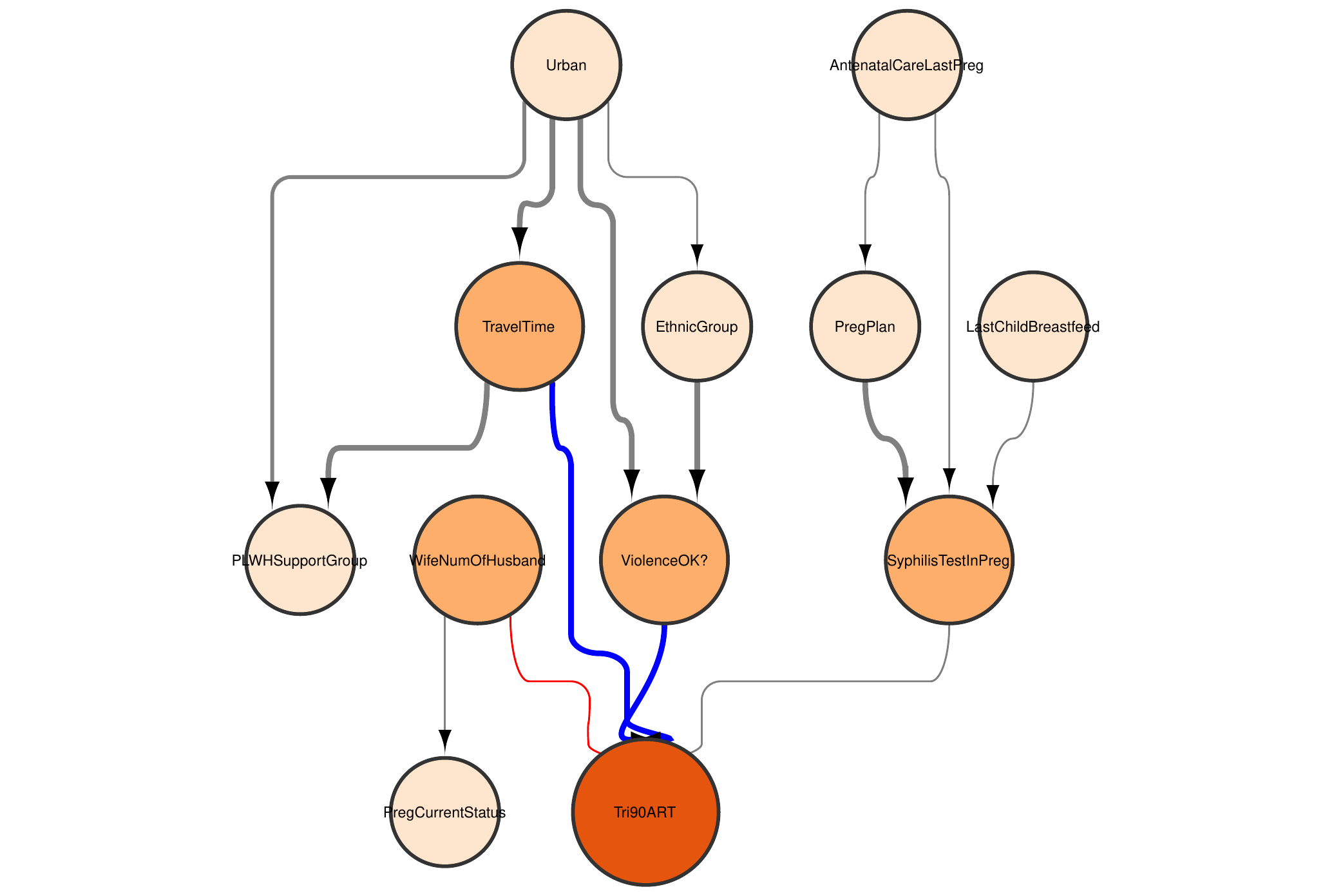"}
    \end{adjustbox}
        \captionsetup{width=9in}
        \caption{90-90-90 ART graph in female.
        Vertices representing the Tri90 goals are biggest and marked by orange; vertices closer to goals have bigger sizes and darker colors than those farther away from goals. Widths of edges reflect the significance of the non-directional connection (conditional dependence) between vertices.
        Red and blue edges represent positive and negative relationships with Tri90 goals, respectively.
        Codebook can be found in Supplement~\ref{Codebook:tri90artfemale}.}
        \label{fig:tri90artfemale}
    \end{center}
\end{figure}

\begin{figure}[!ht]
    \begin{center}
        \begin{adjustbox}{center}
        \includegraphics[width=11in,height=7.5in]{"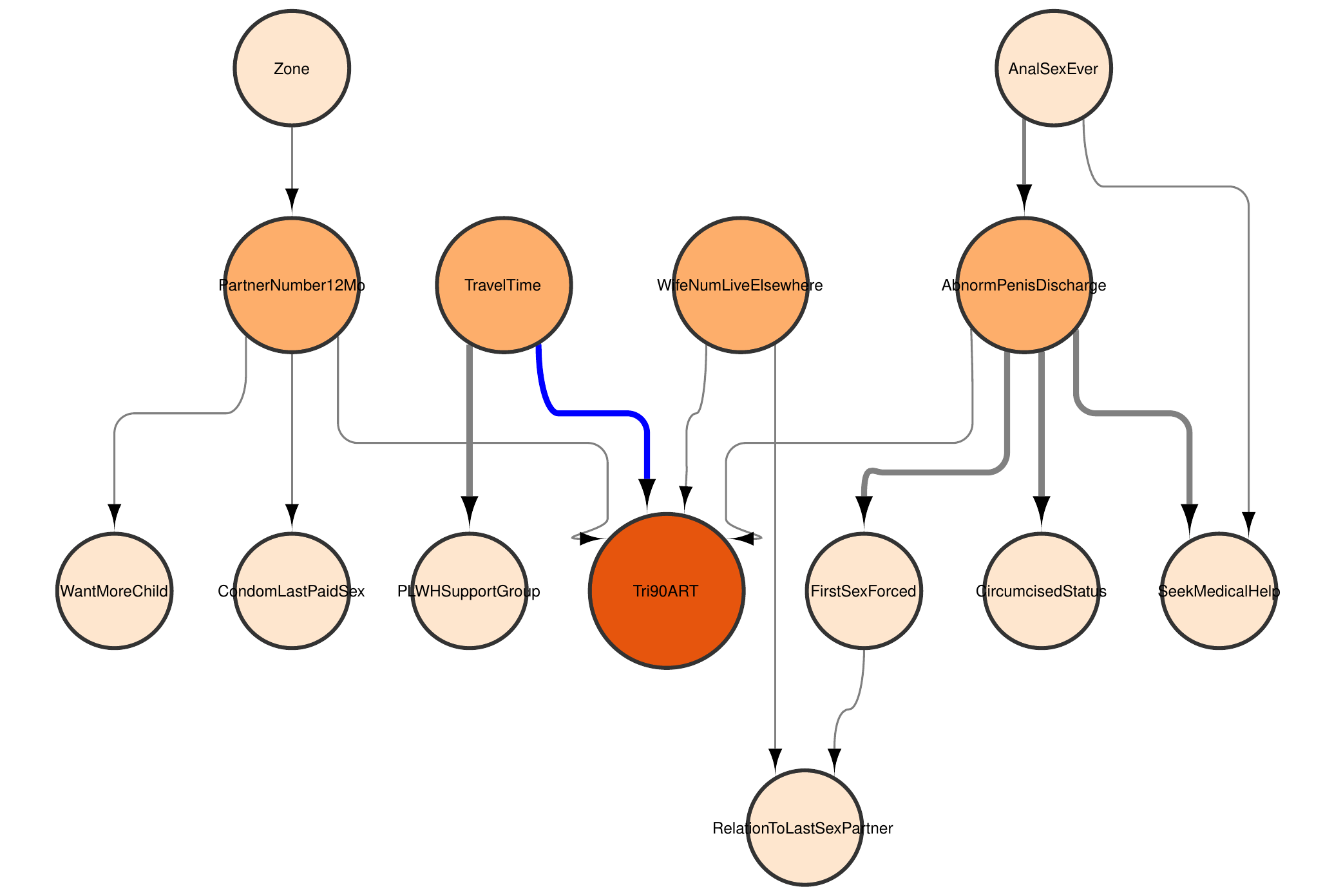"}
        \end{adjustbox}
        \captionsetup{width=9in}
        \caption{90-90-90 ART graph in male.
        Vertices representing the Tri90 goals are biggest and marked by orange; vertices closer to goals have bigger sizes and darker colors than those farther away from goals. Widths of edges reflect the significance of the non-directional connection (conditional dependence) between vertices.
        Red and blue edges represent positive and negative relationships with Tri90 goals, respectively.
        Codebook can be found in Supplement~\ref{Codebook:tri90artmale}.}
        \label{fig:tri90artmale}
    \end{center}
\end{figure}

\begin{figure}[!ht]
    \begin{center}
        \begin{adjustbox}{center}
        \includegraphics[width=11in,height=7.5in]{"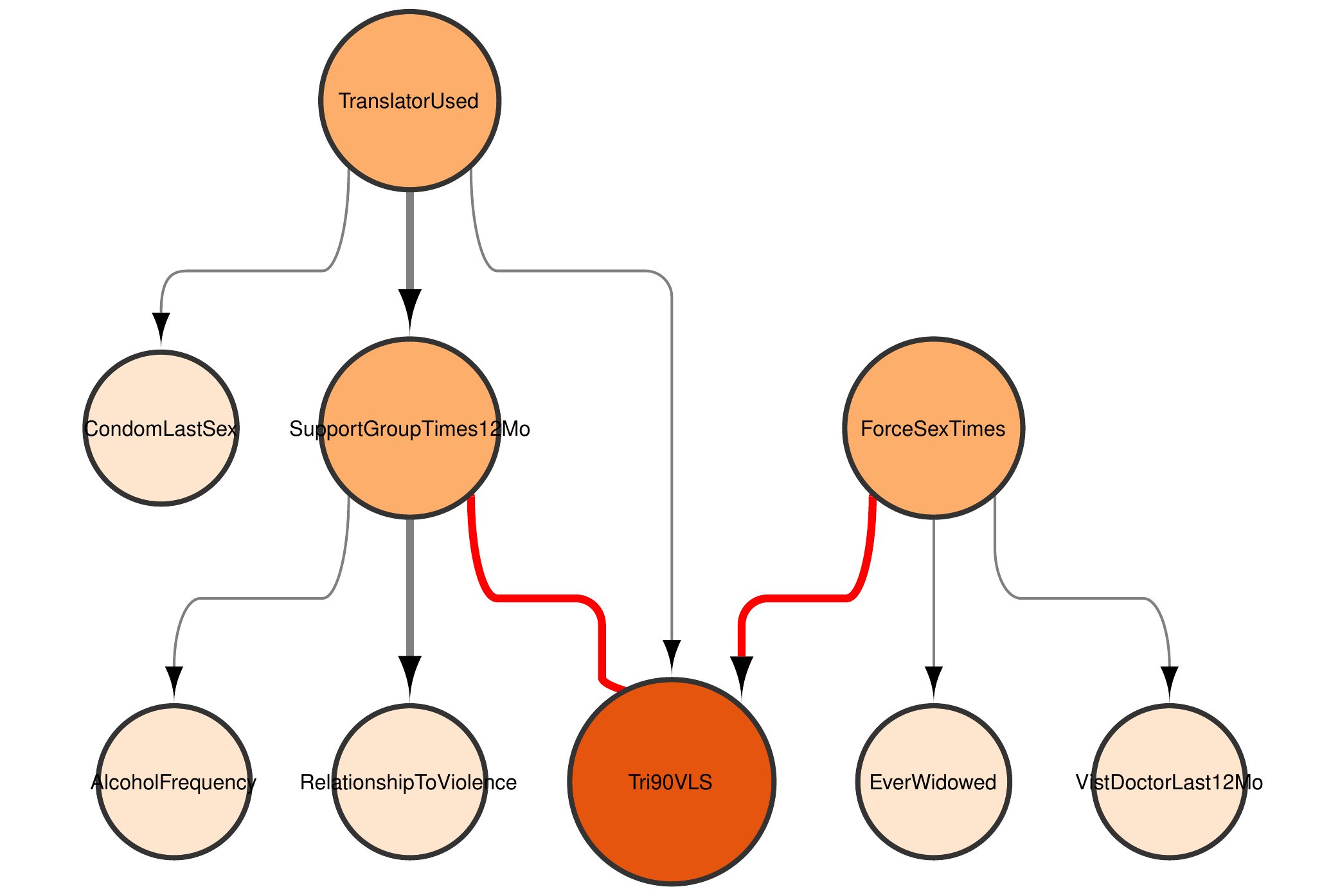"}
        \end{adjustbox}
        \captionsetup{width=9in}
        \caption{90-90-90 VLS graph in female.
        Vertices representing the Tri90 goals are biggest and marked by orange; vertices closer to goals have bigger sizes and darker colors than those farther away from goals. Widths of edges reflect the significance of the non-directional connection (conditional dependence) between vertices.
        Red and blue edges represent positive and negative relationships with Tri90 goals, respectively.
         Codebook can be found in Supplement~\ref{Codebook:tri90vlsfemale}.}
        \label{fig:tri90vlsfemale}
    \end{center}
\end{figure}

\begin{figure}[!ht]
    \begin{center}
        \begin{adjustbox}{center}
        \includegraphics[width=11in,height=7.5in]{"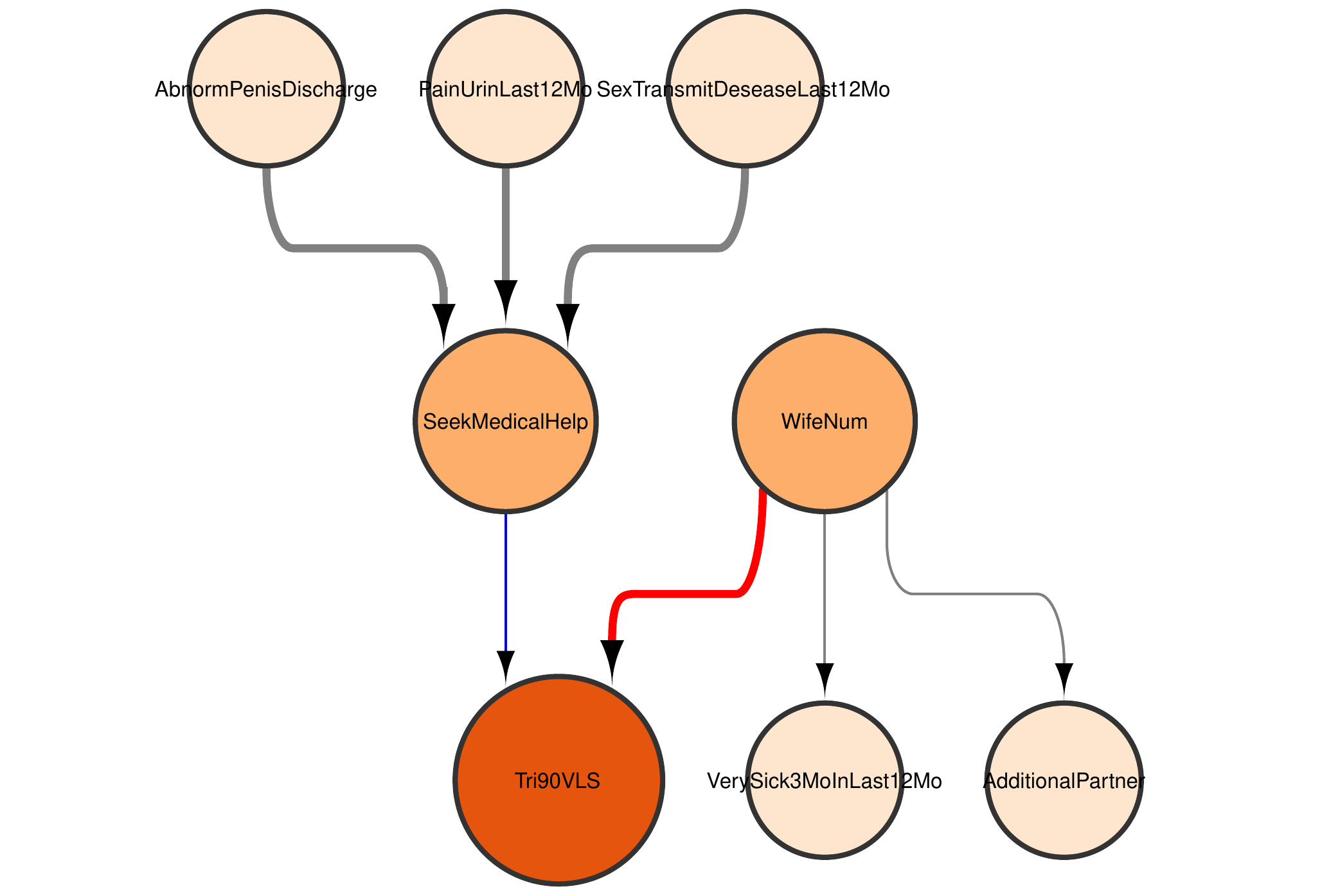"}
        \end{adjustbox}
        \captionsetup{width=9in}
        \caption{90-90-90 VLS graph in male.
        Vertices representing the Tri90 goals are biggest and marked by orange; vertices closer to goals have bigger sizes and darker colors than those farther away from goals. Widths of edges reflect the significance of the non-directional connection (conditional dependence) between vertices.
        Red and blue edges represent positive and negative relationships with Tri90 goals, respectively.
        Codebook can be found in Supplement~\ref{Codebook:tri90vlsmale}.}
        \label{fig:tri90vlsmale}
    \end{center}
\end{figure}

\newpage
\clearpage

\KOMAoptions{paper=a4}
\recalctypearea
\restoregeometry

\section{Additional Simulation Studies}
\label{section:AdditionalSimulation}

\subsection{Chosen of $M_{\text{CI}}$}
\label{section:MCI}

In this simulation study, we use a simulation setting similar to Section~\ref{section:Simulations} to check the performance of the proposed algorithm with different values of $M_{\text{CI}}$.
More specifically, we use the DAGs learned by the proposed algorithm as the truth to generate the simulation data.
That is to say, let $\mathcal{G}_k$ be the DAG learned by the proposed algorithm on the 90-90-90 MPHIA data set $\tf{D}_k$ for $k=1,2,\cdots,6$.
Then we fit the data distribution $\mathcal{P}_k$ based on $\mathcal{G}_k$ on the data $\tf{D}_k$.
We further randomly generate $M$ simulated data sets $\mathcal{D}_k = (\tf{D}_{k,1},\tf{D}_{k,2},\cdots,\tf{D}_{k,M})$ based on the distribution $\mathcal{P}_k$ with the sample size $n$.
Here we set $n = 500$ for a sample size similar to the MPHIA datasets.
And we further apply the proposed algorithm with $M_{\text{CI}} = 2,3,4,5,\infty$ on the generated dataset.
The whole simulation is repeated $500$ times,
and we summarize the empirical averages of true positive rates and true negative rates of edges disregarding the orientation in Table~\ref{tab:MCI}.


\begin{table}

\caption{\label{tab:MCI}Empirical true positive rates and true negative rates of the proposed algorithm with different $M_{\text{CI}}$ (in percentage). Aware, ART, and VLS stand for the three 90-90-90 targets of HIV awareness, ART treatment, and viral load suppression respectively.}
\centering
\resizebox{\linewidth}{!}{
\begin{tabular}[t]{llrrrrrrrrrr}
\toprule
\multicolumn{2}{c}{ } & \multicolumn{5}{c}{True Positive Rate} & \multicolumn{5}{c}{True Negative Rate} \\
\cmidrule(l{3pt}r{3pt}){3-7} \cmidrule(l{3pt}r{3pt}){8-12}
Goals & Gender & $M_{\text{CI}}=2$ & 3 & 4 & 5 & $\infty$ & $M_{\text{CI}}=2$ & 3 & 4 & 5 & $\infty$\\
\midrule
 & Male & 41.4 & 43.1 & 43.2 & 43.3 & 43.3 & 98.1 & 98.0 & 98.0 & 98.0 & 98.0\\

\multirow{-2}{*}{\raggedright\arraybackslash Aware} & Female & 38.4 & 38.2 & 38.3 & 38.4 & 38.4 & 98.6 & 98.4 & 98.4 & 98.4 & 98.4\\
\cmidrule{1-12}
 & Male & 45.2 & 46.4 & 46.4 & 46.4 & 46.4 & 98.2 & 98.0 & 98.0 & 97.9 & 97.9\\

\multirow{-2}{*}{\raggedright\arraybackslash ART} & Female & 37.7 & 37.5 & 37.6 & 37.8 & 38.2 & 98.3 & 98.1 & 98.1 & 98.0 & 98.0\\
\cmidrule{1-12}
 & Male & 41.6 & 44.7 & 44.7 & 44.7 & 44.7 & 97.9 & 97.9 & 97.9 & 97.9 & 97.9\\

\multirow{-2}{*}{\raggedright\arraybackslash VLS} & Female & 40.4 & 40.0 & 40.0 & 40.1 & 40.1 & 98.5 & 98.4 & 98.4 & 98.4 & 98.3\\
\bottomrule
\end{tabular}}
\end{table}

The left and right panels of Table~\ref{tab:MCI} summarize the empirical true positive and negative rates of the proposed algorithm with different $M_{\text{CI}}$, respectively.
From Table~\ref{tab:MCI}, we can see that there is no significant difference among the true positive rates and true negative rates for the proposed algorithm with $M_{\text{CI}}=2,3,4,5,\infty$. 
It shows that the proposed algorithm is quite robust to the choice of $M_{\text{CI}}$.
As discussed by other causal structural learning literature such as \citet{yan2020effective}, we recommend to use $M_{\text{CI}} = 3$ for sparse or moderate sparse graphs.
Notice that the choice of value of $M_{\text{CI}}$ depends on the sample size, the types of covariates, the property of the true graph (the degree of the graph), etc as discussed in the remark for Algorithm~\ref{alg:step1}, it is also possible to try different values of $M_{\text{CI}}$ and to cross-validation methods, BIC scores, or simulation studies to have a more sophisticated chosen of $M_{\text{CI}}$.

\subsection{Simulation Study to Check Receiver Operating Characteristic (ROC) Curve}
\label{section:Simulation2}

In this simulation study, we use a simulation setting similar to Section~\ref{section:Simulations}. We use the receiver operating characteristic (ROC) curve and the area under the (ROC) curve (AUC) to have a closer look at the performance of the different structural learning algorithms.
More specifically, we use the DAGs learned by the proposed algorithm as the truth to generate the simulation data since
the BIC criterion shows that the graphs learned by the proposed algorithm are better fits for the MPHIA data than those learned by the other algorithms.
That is to say, let $\mathcal{G}_k$ be the DAG learned by the proposed algorithm on the 90-90-90 MPHIA data set $\tf{D}_k$ for $k=1,2,\cdots,6$.
Then we fit the data distribution $\mathcal{P}_k$ based on $\mathcal{G}_k$ on the data $\tf{D}_k$.
We further randomly generate $M$ simulated data sets $\mathcal{D}_k = (\tf{D}_{k,1},\tf{D}_{k,2},\cdots,\tf{D}_{k,M})$ based on the distribution $\mathcal{P}_k$ with the same sample size as the original data set $\tf{D}_k$.
Applying graphical learning algorithm $A_i$, for $i=1,\cdots,5$, on the simulated data sets $\mathcal{D}_k$, we have $M$ DAGs $(\mathcal{G}_{k,i,1},\mathcal{G}_{k,i,2},\cdots,\mathcal{G}_{k,i,M})$.
We set $M = 500$, so the whole simulation is repeated $500$ times.

\begin{figure}[!ht]
    \centering
    \begin{subfigure}[t]{0.49\textwidth}
        \centering
        \includegraphics[width=\textwidth]{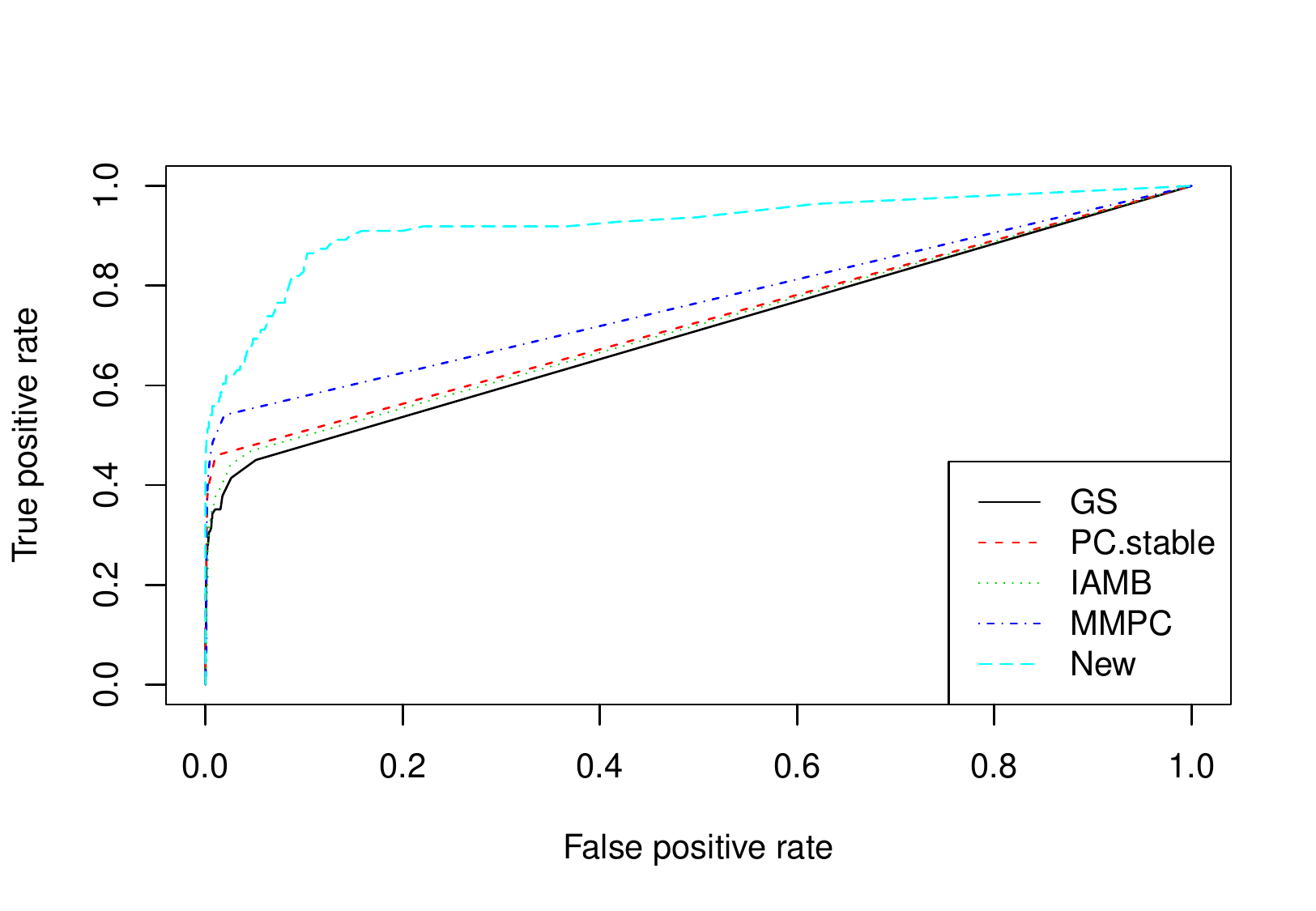}
        \caption{\label{fig:ROCawareMale} ROC for male awareness.}
    \end{subfigure}
    \begin{subfigure}[t]{0.49\textwidth}
        \centering
        \includegraphics[width=\textwidth]{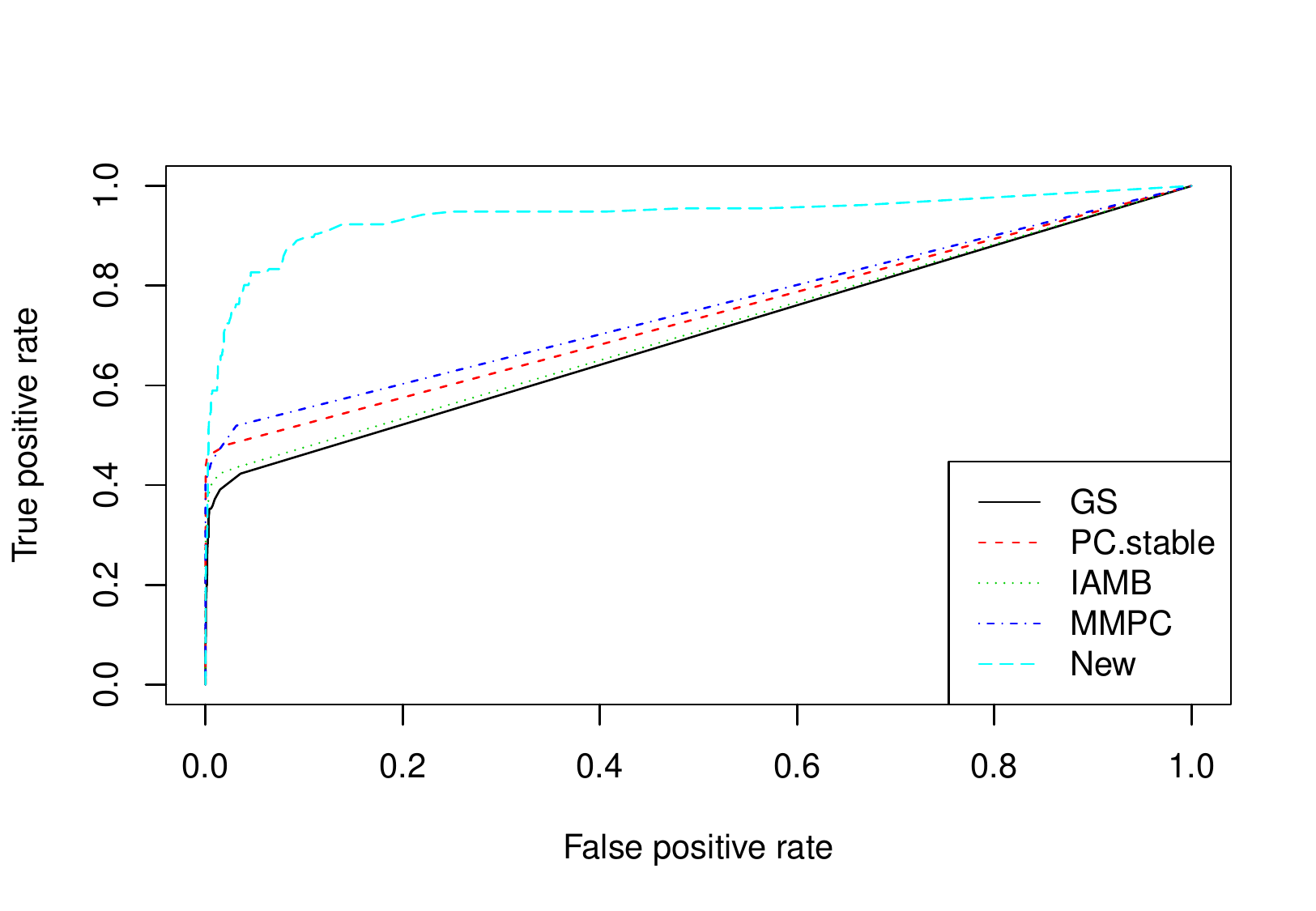}
        \caption{\label{fig:ROCawareFemale} ROC for female awareness.}
    \end{subfigure}
    \begin{subfigure}[t]{0.49\textwidth}
        \centering
        \includegraphics[width=\textwidth]{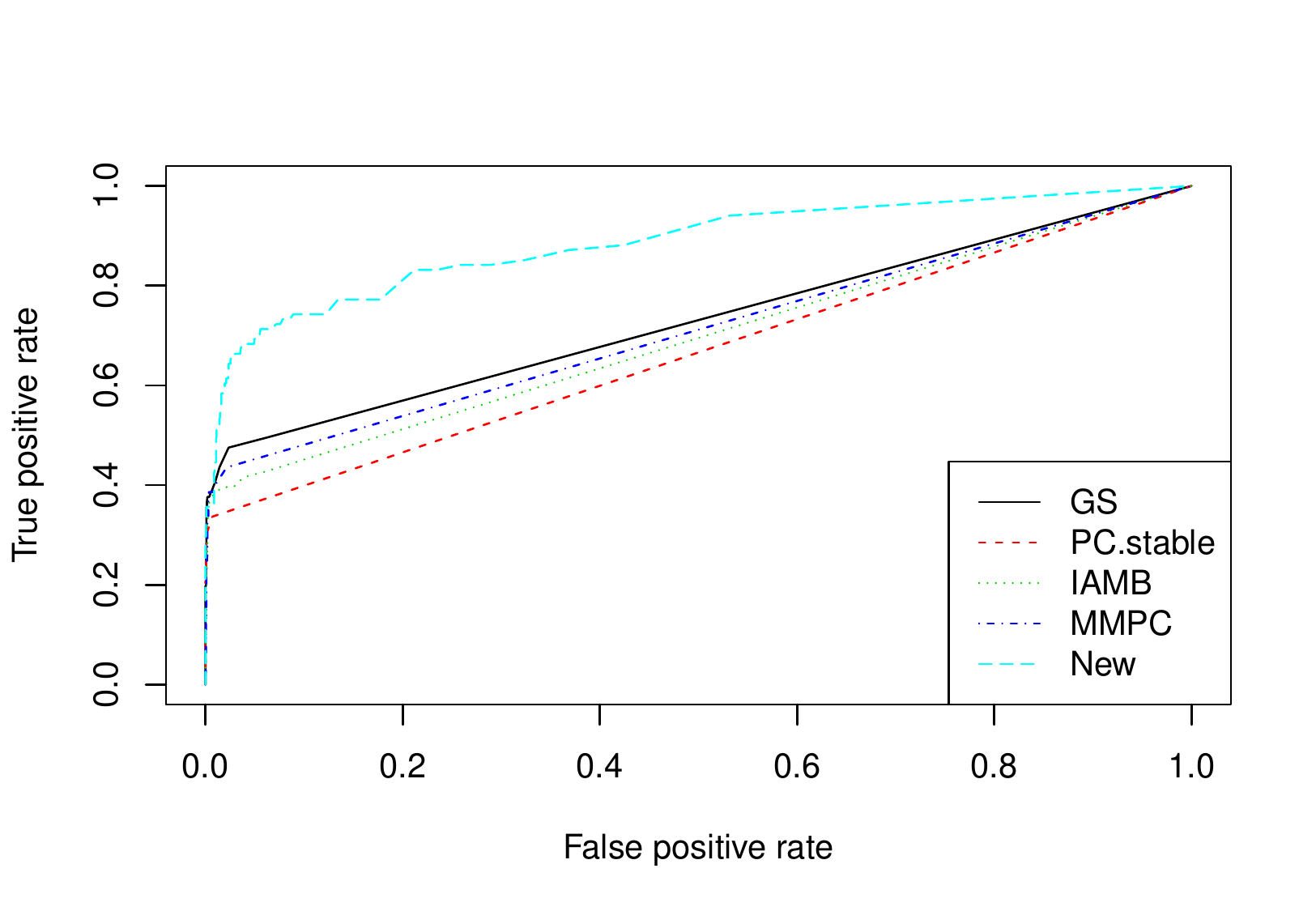}
        \caption{\label{fig:ROCartMale} ROC for male ART.}
    \end{subfigure}
    \begin{subfigure}[t]{0.49\textwidth}
        \centering
        \includegraphics[width=\textwidth]{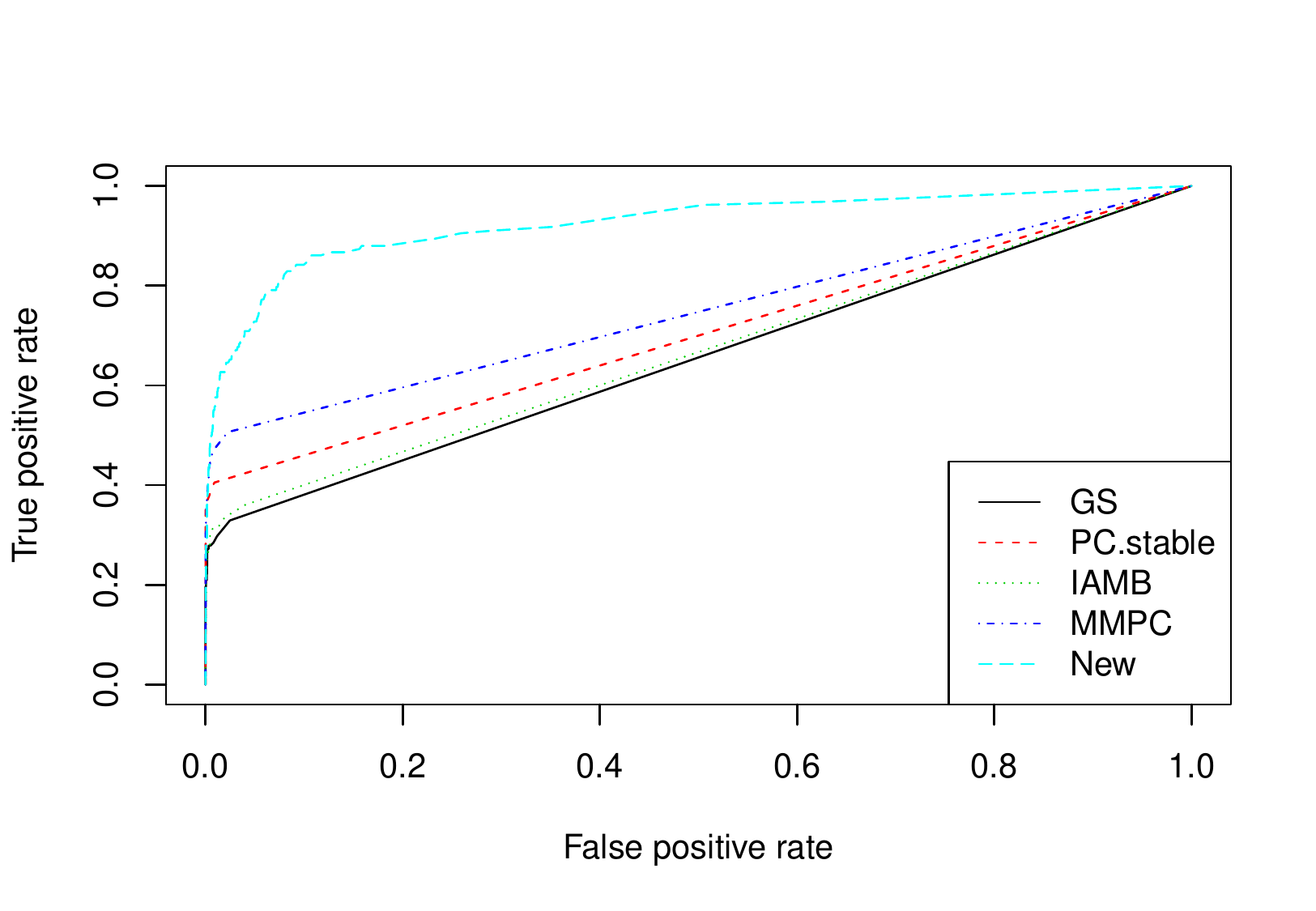}
        \caption{\label{fig:ROCartFemale} ROC for female ART.}
    \end{subfigure}
    \begin{subfigure}[t]{0.49\textwidth}
        \centering
        \includegraphics[width=\textwidth]{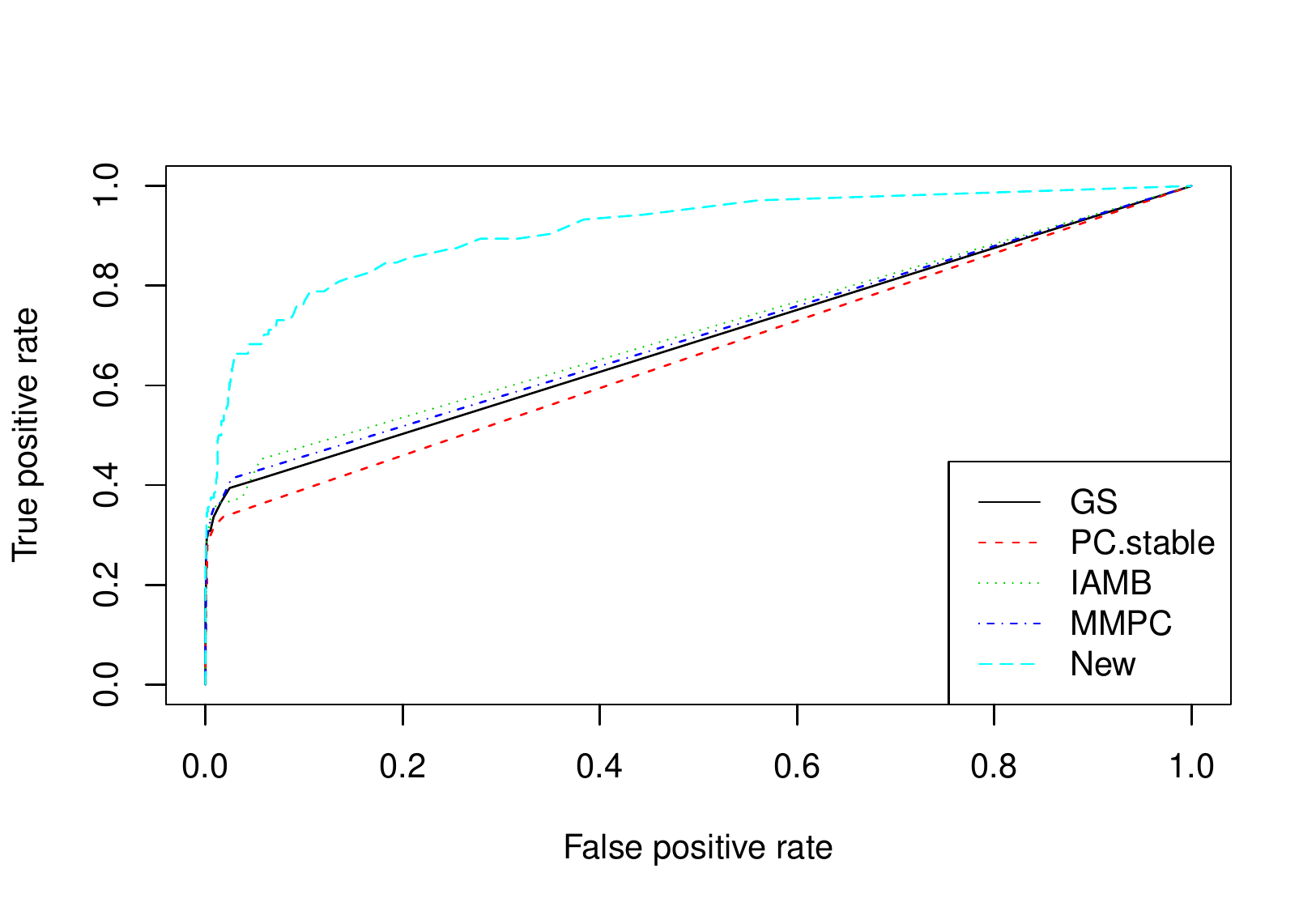}
        \caption{\label{fig:ROCvlsMale} ROC for male VLS.}
    \end{subfigure}
    \begin{subfigure}[t]{0.49\textwidth}
        \centering
        \includegraphics[width=\textwidth]{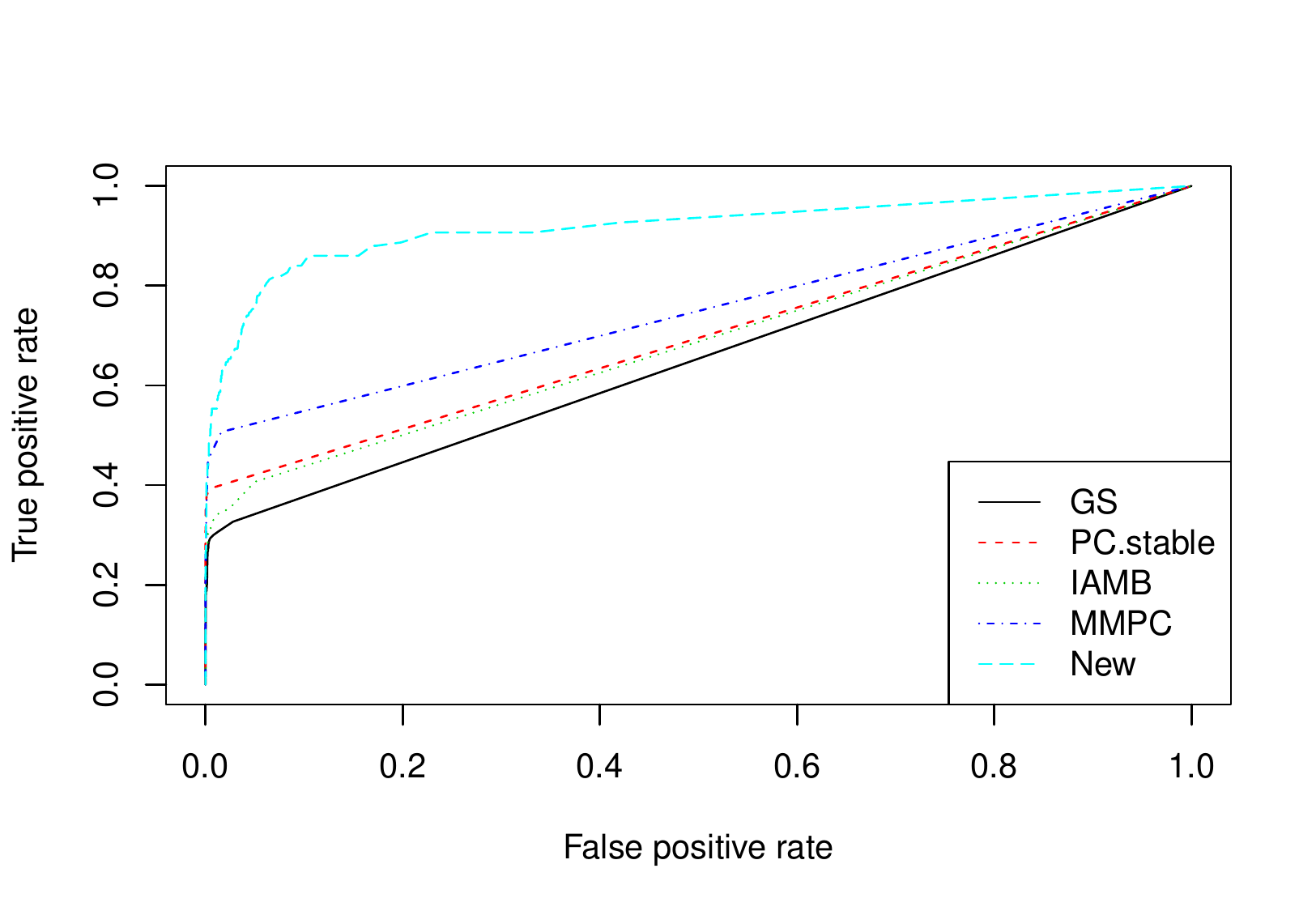}
        \caption{\label{fig:ROCvlsFemale} ROC for female VLS.}
    \end{subfigure}
    \begin{singlespace}
    \caption{ROC curves for edge discovery of different structural learning algorithms for three 90-90-90 targets of both genders calculated from 500 Monte Carlo simulations.}
    \label{fig:ROC}
    \end{singlespace}
\end{figure}

\begin{figure}[!ht]
    \centering
    \begin{subfigure}[t]{0.49\textwidth}
        \centering
        \includegraphics[width=\textwidth]{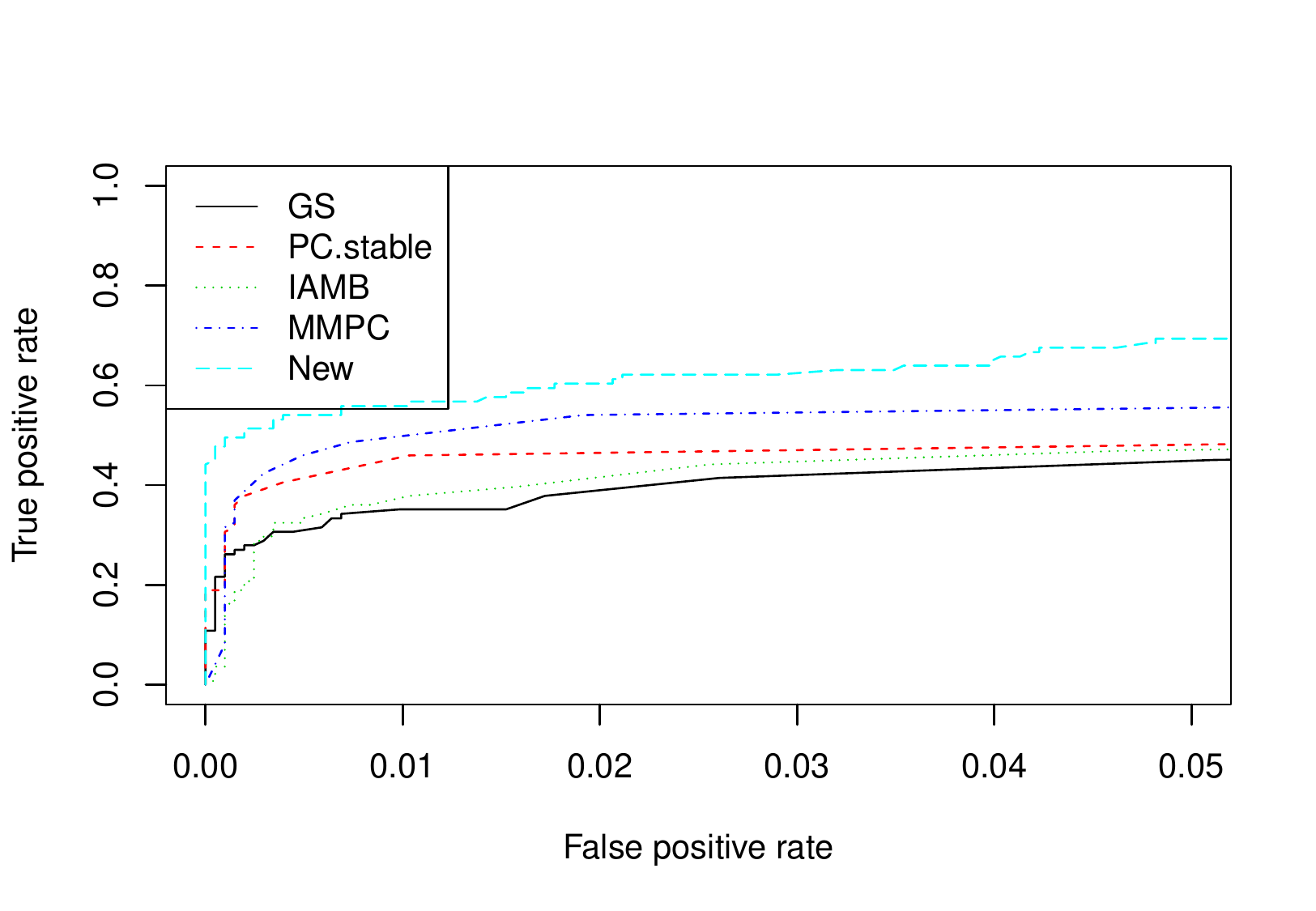}
        \caption{\label{fig:ROCawareMale2}  Part of ROC for male awareness.}
    \end{subfigure}
    \begin{subfigure}[t]{0.49\textwidth}
        \centering
        \includegraphics[width=\textwidth]{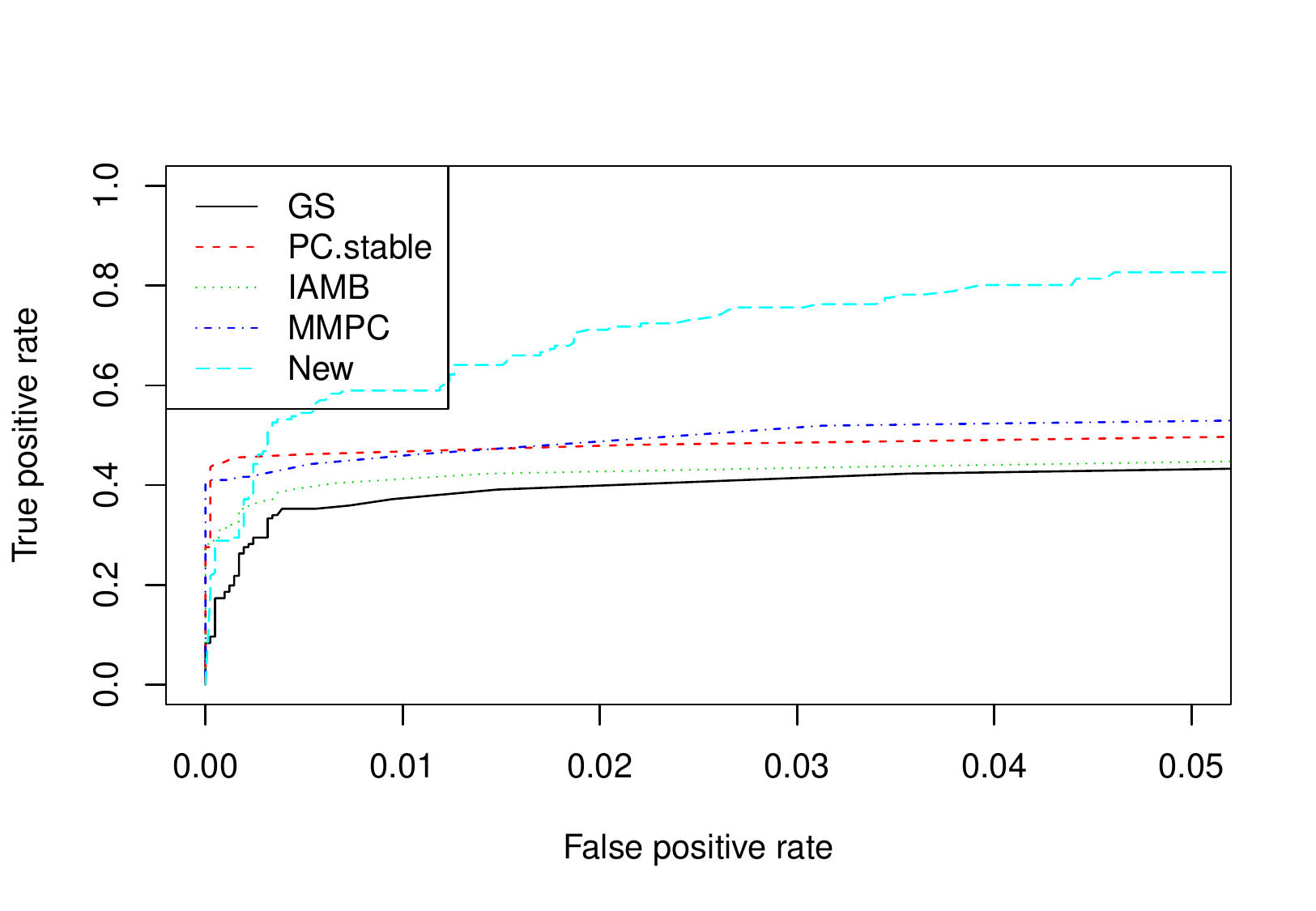}
        \caption{\label{fig:ROCawareFemale2}  Part of ROC for female awareness.}
    \end{subfigure}
    \begin{subfigure}[t]{0.49\textwidth}
        \centering
        \includegraphics[width=\textwidth]{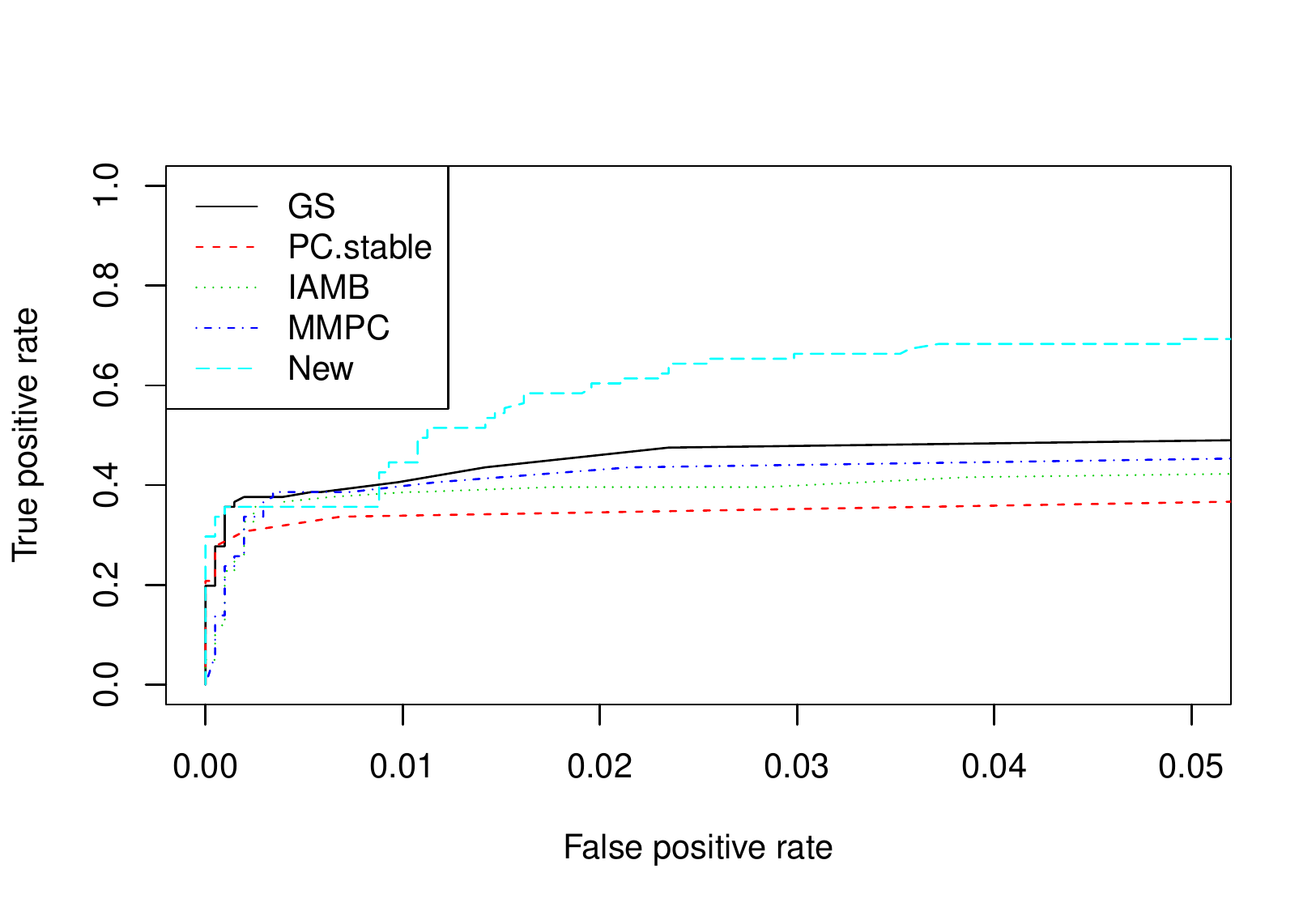}
        \caption{\label{fig:ROCartMale2}  Part of ROC for male ART.}
    \end{subfigure}
    \begin{subfigure}[t]{0.49\textwidth}
        \centering
        \includegraphics[width=\textwidth]{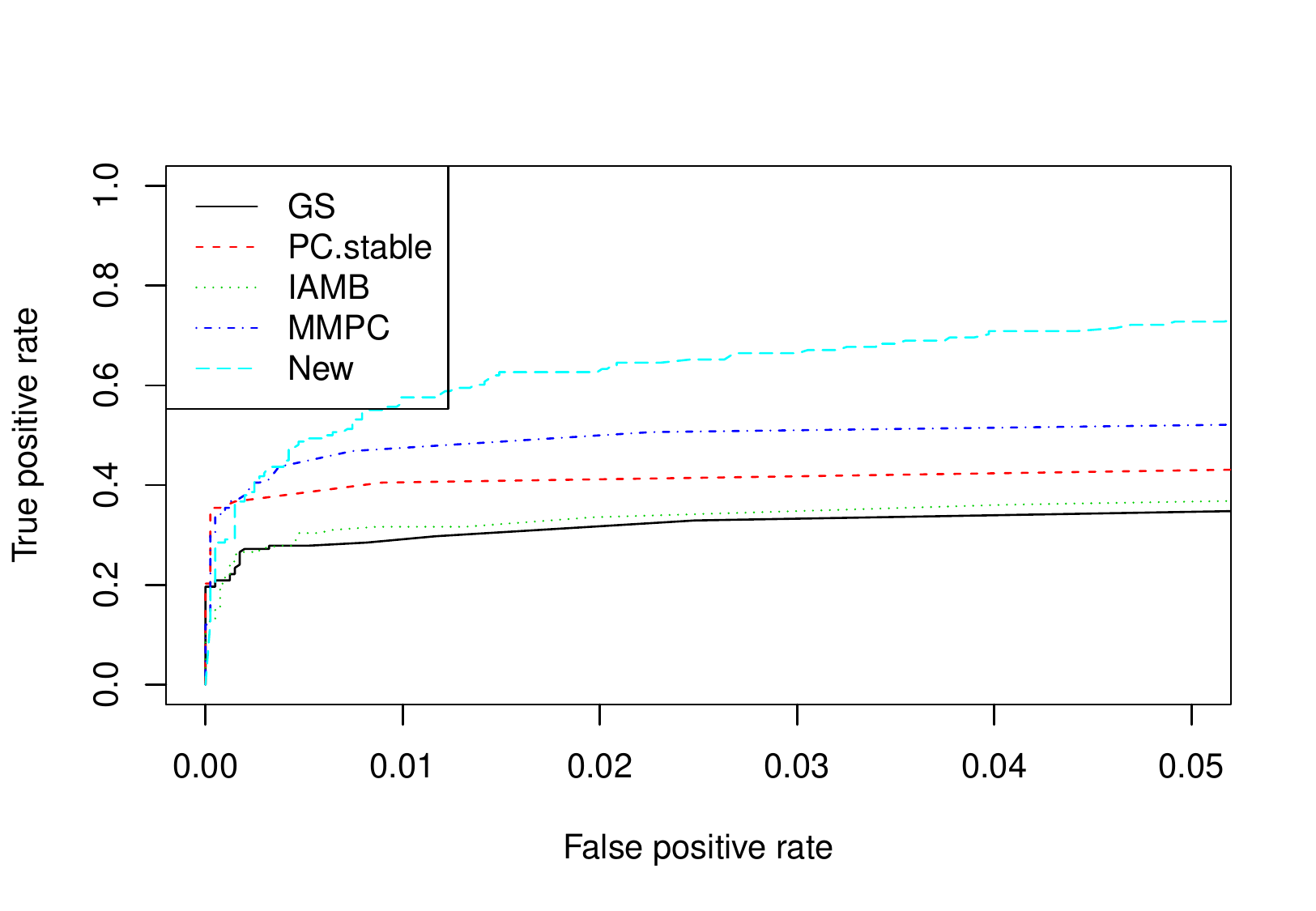}
        \caption{\label{fig:ROCartFemale2}  Part of ROC for female ART.}
    \end{subfigure}
    \begin{subfigure}[t]{0.49\textwidth}
        \centering
        \includegraphics[width=\textwidth]{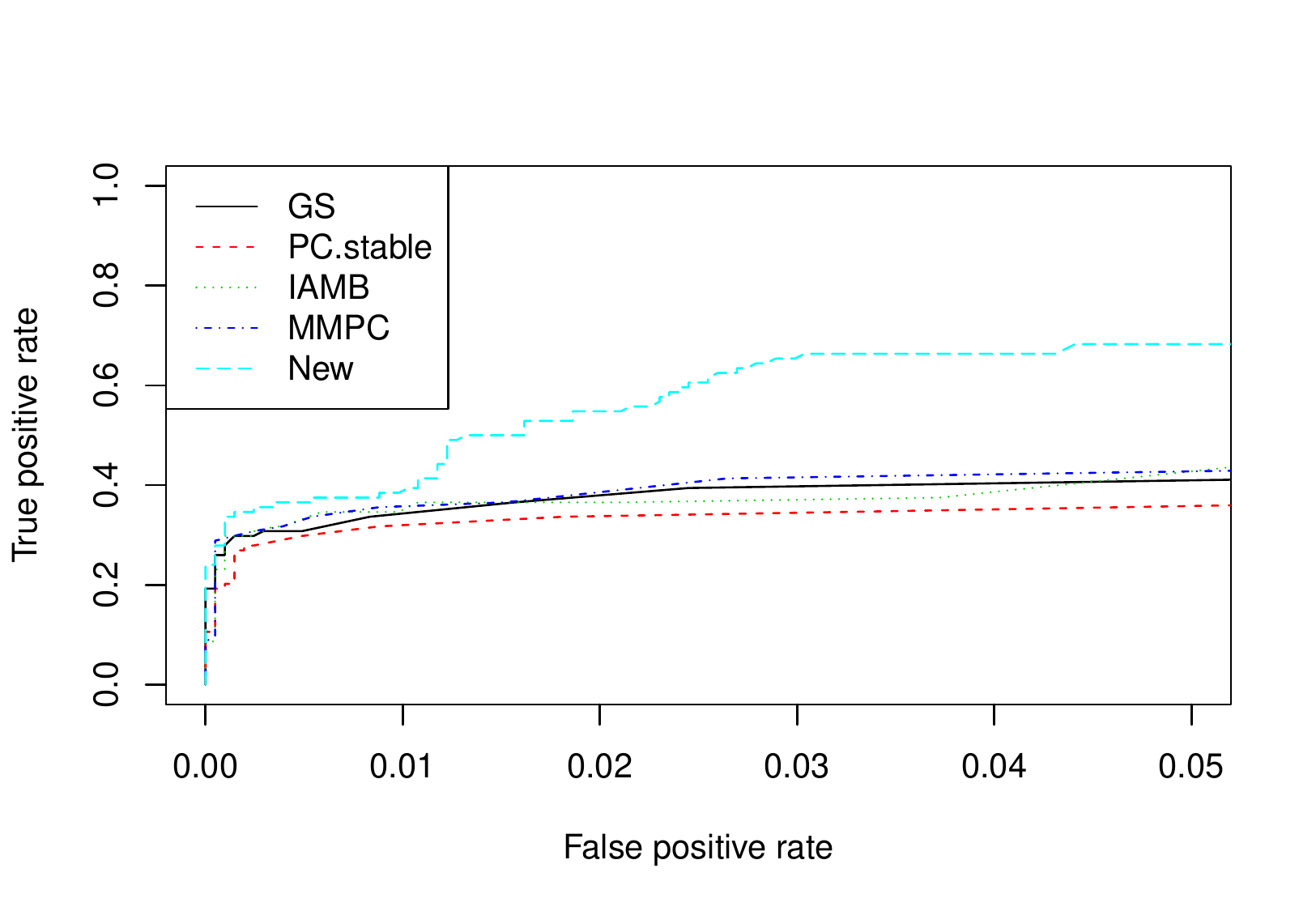}
        \caption{\label{fig:ROCvlsMale2}  Part of ROC for male VLS.}
    \end{subfigure}
    \begin{subfigure}[t]{0.49\textwidth}
        \centering
        \includegraphics[width=\textwidth]{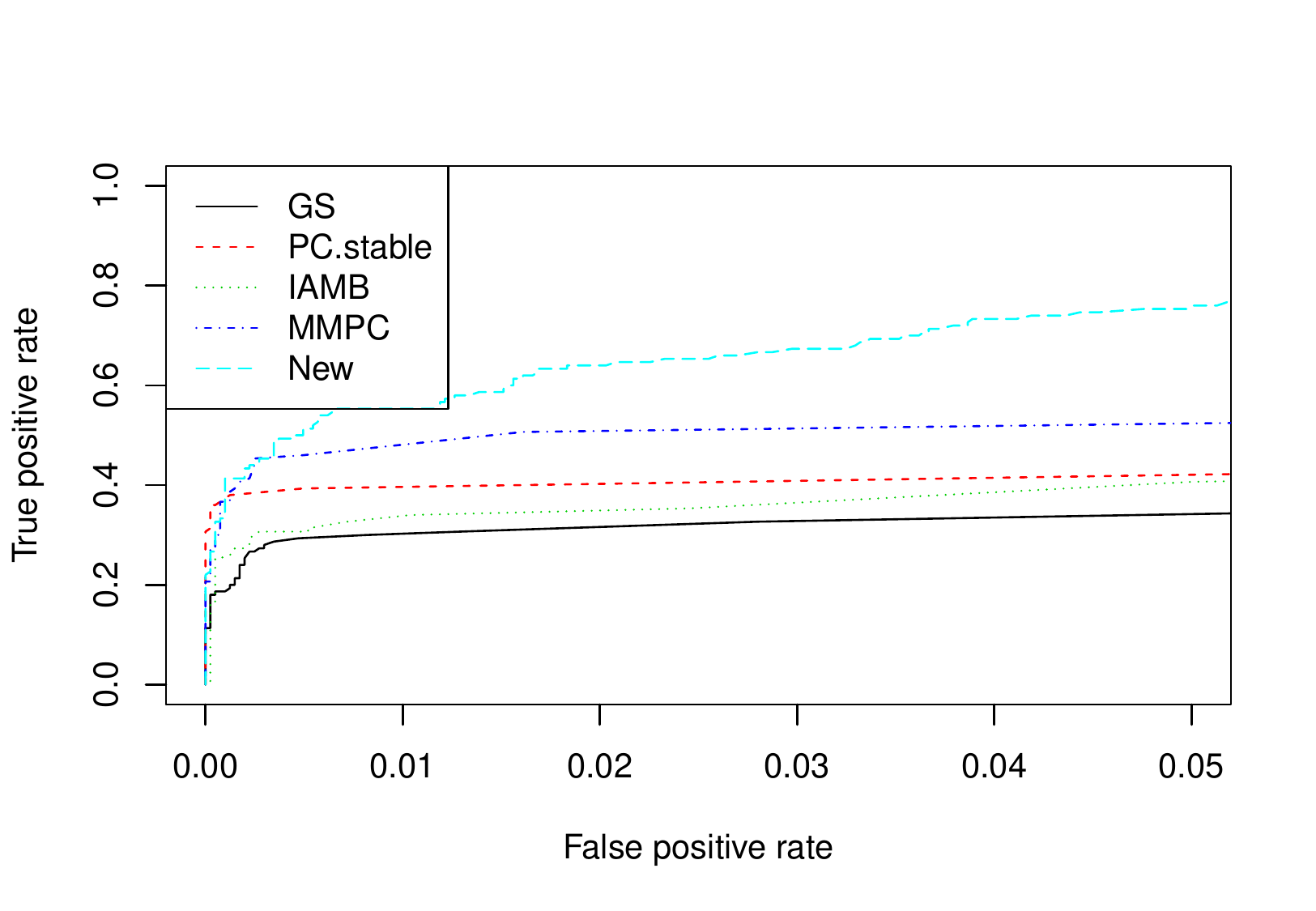}
        \caption{\label{fig:ROCvlsFemale2} Part of ROC for female VLS.}
    \end{subfigure}
    \begin{singlespace}
    \caption{Part of ROC curves with FPR $\leq 0.05$ for edge discovery of different structural learning algorithms for three 90-90-90 targets of both genders calculated from 500 Monte Carlo simulations.}
    \label{fig:ROCsmall}
    \end{singlespace}
\end{figure}

Here we use the receiver operating characteristic (ROC) curve and the area under the (ROC) curve (AUC) to measure the edge discovery in 90-90-90 graphs to have a better understanding of the Type I and Type II error and their trade-off for different structural learning algorithms.
To get the ROC curve,
from the $M$ DAGs learned in the Monte Carlo simulations $(\mathcal{G}_{k,i,1},\mathcal{G}_{k,i,2},\cdots,\mathcal{G}_{k,i,M})$,
we first calculate an average graph $\bar{\mathcal{G}}_{k,i}$.
$\bar{\mathcal{G}}_{k,i}$ is an undirected graph with weighted edges where the weight of an edge $X - Y$ is the empirical frequency of the existence of edge $X - Y$ in $(\mathcal{G}_{k,i,1},\mathcal{G}_{k,i,2},\cdots,\mathcal{G}_{k,i,M})$ disregarding the direction of edges.
Hence $\bar{\mathcal{G}}_{k,i}$ reflects the ``confidence" in edges for algorithm $A_i$.
Then for each cut-off value $\lambda$, we can get an undirected graph $\mathcal{G}_{k,i,\lambda}$ by keeping all the edges in $\bar{\mathcal{G}}_{k,i}$ with weights greater than or equal to $\lambda$,
calculate the true positive rate (TPR) and false positive rate (FPR) of edges,
and obtain the AUC score and the ROC curve.
The AUC score and the ROC curve for edge discovery in 90-90-90 graphs of different structural learning algorithms calculated from 500 Monte Carlo simulations are summarized in Table~\ref{tab:simuAUC} and Figures~\ref{fig:ROC}, \ref{fig:ROCsmall}, respectively.

In Table~\ref{tab:simuAUC}, we can see that the proposed algorithm achieves better (larger) AUC compared to other structural learning algorithms across all the three 90-90-90 goals and both genders.
To understand why the proposed method achieves a better AUC, let us look at Figures~\ref{fig:ROC} and \ref{fig:ROCsmall}.
In Figure~\ref{fig:ROC}, we can see that when the false positive rate (FPR) is extremely small, the proposed algorithm has a similar true positive rate (TPR) as the other algorithms;
while with larger FPR, the proposed algorithm has better TPR than the other algorithms.
We can see the details for the ROCs with small FPR more clearly in Figure~\ref{fig:ROCsmall} and find that the proposed algorithm achieves a better or comparable TPR with FPR $\geq 0.01$ and a much better TPR with FPR $\geq 0.02$ across all 90-90-90 goals and genders.
The similar TPR of all algorithms for extremely small FPR illustrates that all algorithms have similar performance in the discovery of the most important relationship from the data.
Furthermore, the better TPR of the proposed algorithm for larger FPR shows that
while the existing structural learning algorithms cannot discover weaker signals beyond a cut-point; the proposed algorithm has a better ability in picking up relatively weak signals,
which is the reason for the better AUC of the proposed algorithm in Table~\ref{tab:simuAUC}.

\begin{singlespace}

\begin{table}

\caption{\label{tab:simuAUC}AUC of Different 90-90-90 Targets, Genders, and Causal Structural Learning Algorithms. Aware, ART, and VLS stand for the three 90-90-90 targets of HIV awareness, ART treatment, and viral load suppression respectively.
}
\centering
\begin{tabular}[t]{lrrrrrr}
\toprule
\multicolumn{1}{c}{ } & \multicolumn{2}{c}{Aware} & \multicolumn{2}{c}{ART} & \multicolumn{2}{c}{VLS} \\
\cmidrule(l{3pt}r{3pt}){2-3} \cmidrule(l{3pt}r{3pt}){4-5} \cmidrule(l{3pt}r{3pt}){6-7}
Method & Male & Female & Male & Female & Male & Female\\
\midrule
PC-stable & 0.726 & 0.734 & 0.666 & 0.700 & 0.662 & 0.695\\
GS & 0.708 & 0.700 & 0.730 & 0.655 & 0.688 & 0.653\\
IAMB & 0.719 & 0.708 & 0.694 & 0.666 & 0.706 & 0.686\\
MMPC & 0.764 & 0.751 & 0.711 & 0.747 & 0.697 & 0.749\\
New & 0.922 & 0.940 & 0.885 & 0.925 & 0.910 & 0.918\\
\bottomrule
\end{tabular}
\end{table}

\end{singlespace}

\textcolor{black}{\subsection{Simulation Study with Continuous Variables and Different Graphical Densities and Signal Strengths}}

In this simulation study, we check the empirical performance of the proposed algorithm on synthetic data sets with continuous variables and different levels of ``sparsity" and signal strengths of edges.
Let $K$ be the number of vertices and $\rho \in (0, 1)$ be the parameter that controls the level of ``sparsity" of edges, we generate the simulation data randomly using the following procedure:
\begin{enumerate}
\item
We first generate a DAG $\mathcal{G}^{*}$.
Generate $K(K-1)/2$ random variables $E^{*}_{i,j}$ i.i.d from $\operatorname{Bernoulli}(\rho)$, $1 \leq i < j \leq K$.
For vertices $i$ and $j$, the edge $i \rightarrow j$ exists in $\mathcal{G}^{*}$ if and only if $E^{*}_{i,j} = 1$.
Further generate $K(K-1)/2$ random variables $S^{*}_{i,j}$ i.i.d from $\operatorname{Normal}(0, 1)$, $1 \leq i < j \leq K$.

\item 
We then generate a dataset $\tf{D}^{*}$ of size $n$ according to DAG $\mathcal{G}^{*}$.
For $j=1,\cdots,K$, generate $x_{j}^{*}$ recursively from the following linear regression models:
\begin{equation}
x_{j}^{*} = \theta \sum_{i=1}^{j-1} x_{i}^{*} E^{*}_{i,j} S^{*}_{i,j} + \epsilon_{j}^{*},
\end{equation}
where $\theta$ controls the strengths of signals and $\epsilon_{j}^{*}$ i.i.d. follows the standard normal distribution, $j=1,\cdots,K$.
And we repeat this step $n$ times to generate an $n$ by $K$ data set $\tf{D}^{*}$.

\end{enumerate}

After generation of the dataset $\tf{D}^{*}$, we permute the order of variables and use the permutation to obtain an $n$ by $K$ data set $\tf{D}$ and the corresponding DAG $\mathcal{G}$.
We then carry out the proposed algorithm together with the aforementioned PC-stable, GS, MMPC, and IAMB algorithms on the $n$ by $K$ data set $\tf{D}$.
Furthermore, we calculate true positive rates and true negative rates of edges disregarding the orientation for each algorithm.

\begin{singlespace}

\begin{table}[!ht]
\caption{\label{tab:oldsimulation}Empirical true positive rates and true negative rates of different causal structural learning algorithms (in percentage).
}
\resizebox{\linewidth}{!}{
\centering
\begin{tabular}[t]{rrrrrrrrrrrr}
\toprule
\multicolumn{2}{c}{ } & \multicolumn{5}{c}{True Positive Rate} & \multicolumn{5}{c}{True Negative Rate} \\
\cmidrule(l{3pt}r{3pt}){3-7} \cmidrule(l{3pt}r{3pt}){8-12}
$\rho$ & $\theta$ & GS & PC-stable & IAMB & MMPC & New & GS & PC-stable & IAMB & MMPC & New\\
\midrule
 & 0.125 & 32.30 & 35.6 & 35.4 & 35.5 & 37.8 & 99.3 & 99.3 & 99.3 & 99.3 & 99.1\\

 & 0.25 & 50.60 & 61.5 & 60.4 & 60.8 & 63.7 & 99.4 & 99.4 & 99.5 & 99.4 & 99.1\\

 & 0.5 & 55.08 & 77.3 & 74.6 & 76.0 & 79.9 & 99.6 & 99.6 & 99.6 & 99.6 & 99.2\\

\multirow{-4}{*}{\raggedleft\arraybackslash 0.01} & 0.75 & 54.14 & 82.7 & 78.8 & 81.3 & 85.3 & 99.6 & 99.7 & 99.7 & 99.7 & 99.3\\
\cmidrule{1-12}
 & 0.125 & 29.29 & 34.8 & 34.0 & 34.1 & 37.5 & 99.4 & 99.4 & 99.4 & 99.4 & 99.1\\

 & 0.25 & 39.84 & 59.3 & 55.0 & 55.8 & 62.6 & 99.6 & 99.6 & 99.7 & 99.6 & 99.2\\

 & 0.5 & 33.44 & 74.2 & 63.4 & 67.9 & 78.4 & 99.7 & 99.8 & 99.8 & 99.8 & 99.4\\

\multirow{-4}{*}{\raggedleft\arraybackslash 0.02} & 0.75 & 26.94 & 76.9 & 63.0 & 70.1 & 81.9 & 99.7 & 99.9 & 99.9 & 99.9 & 99.4\\
\cmidrule{1-12}
 & 0.125 & 23.25 & 33.2 & 30.0 & 30.3 & 36.4 & 99.6 & 99.5 & 99.6 & 99.6 & 99.2\\

 & 0.25 & 22.26 & 56.2 & 42.5 & 44.1 & 60.8 & 99.7 & 99.8 & 99.9 & 99.9 & 99.4\\

 & 0.5 & 9.70 & 66.4 & 41.6 & 48.3 & 73.2 & 99.8 & 99.9 & 100.0 & 100.0 & 99.5\\

\multirow{-4}{*}{\raggedleft\arraybackslash 0.04} & 0.75 & 5.46 & 65.0 & 38.2 & 48.3 & 73.9 & 99.8 & 99.9 & 100.0 & 100.0 & 99.4\\
\bottomrule
\end{tabular}}
\end{table}

\end{singlespace}

Here we set $K = 100$ and $n = 500$ for a similar number of covariates and sample size with our real data.
We set $\rho = (0.01,0.02, 0.04)$ for different levels of ``sparsity" of the true graph
and $\theta = (0.125, 0.25, 0.5, 0.75)$ for different strengths of signals.
Note that in this simulation, we set the upper bound of sizes of conditional sets $M_{\text{CI}} = 2$ and the size of the conditional independence test $\alpha=0.01$ for all the causal structural learning algorithms to reduce the computation time.
We repeat the Monte Carlo simulation 1,000 times for each setting and summarize the results in Table~\ref{tab:oldsimulation}.
The left and right panels of Table~\ref{tab:oldsimulation} summarize the empirical true positive and negative rates of the proposed algorithm as well as those of existing algorithms, respectively.
From the right panel of Table~\ref{tab:oldsimulation}, we can see that the proposed algorithm has similar true negative rates with existing algorithms.
Furthermore, from the left panel of Table~\ref{tab:oldsimulation}, we can see that the proposed algorithm has better true positive rates than existing algorithms.
In sum, we have similar conclusions to those of Section~4.

Comparing the simulation results in Table~\ref{tab:oldsimulation} with those in Table~\ref{tab:newsimulation}, notice that the simulation settings in Section~4 are more challenging than those in this section in terms of the true positive rate.
This is because there are many categorical variables in the simulation in Section~4, while there are only continuous ones in the simulation in this section.
Since the conditional set of categorical variables takes more degrees of freedom away from the conditional independence tests than the continuous ones, categorical variables in the simulation in Section~4 can lead to more Type II errors and more contradictory/inconsistent statistical testing results than the simulation in this section. 
Hence, we can see that the improvement in the true positive rates of the proposed algorithm over the existing ones in Table~\ref{tab:newsimulation} is larger than the improvement in Table~\ref{tab:oldsimulation} in this section.
In sum, we can see that the proposed algorithm is more beneficial in the true positive rate in  the  case  of  categorical  variables.


\section{Parts of MPHIA Codebook}
\label{section:Codebook}

\subsection{Codebook for Covariates in Table~\ref{tab:Neighbors}}
\label{Codebook:neighbors}


\begin{enumerate}
\item AbnormPenisDischarge: During the last 12 months, have you had an abnormal discharge from your penis?
\item AgeGroup: Age groups for population pyramid
\item AlcoholFrequency: How often do you have a drink containing alcohol?
\item EasyGetCondom: If you wanted a condom, would it be easy for you to get one?
\item Education: Level of school respondent ever attended
\item ForceSexTimes: How many times in your life have you been physically forced to have sex?
\item PartnerAge: How old is your partner? Please give your best guess.
\item PartnerNumber12Mo: Number of people they had sex with in the last 12 months
\item PLWHSupportGroup: Have you ever attended a support group for people living with HIV?
\item PregNum: How many times have you been pregnant including a current pregnancy?
\item SeekMedicalHelp: Did you see a doctor, clinical officer or nurse because of these problems?
\item SupportGroupTimes12Mo: In the last 12 months, how many times did you attend a support group?
\item SyphilisTestInPreg: When you were pregnant, were you offered a test for syphilis?
\item TranslatorUsed: whether or translator is used or not.
\item TravelTime: At your last HIV care visit, approximately how long did it take you to travel from your home (or workplace) one way?
\item ViolenceOK?: Do you believe it is right for a man to hit or beat his wife/partner?
\item WifeNum: Altogether, how many wives or partners do you have?
\item WifeNumLiveElsewhere: How many wives/partners do you have who live elsewhere?
\item WifeNumOfHusband: Including yourself, in total, how many wives or live-in partners does your husband or partner have?
\end{enumerate}

\subsection{Codebook for Covariates in Figure~\ref{fig:tri90awarefemale}}
\label{Codebook:tri90awarefemale}


\begin{enumerate}
\item AgeGroup: Age groups for population pyramid
\item ChildNumSince2012: How many children have you given birth to since 2012?
\item CircumcisedHIVRisk: Relationship of circumcision and risk of HIV?
\item EasyGetCondom: If you wanted a condom, would it be easy for you to get one?
\item Education: Level of school respondent ever attended
\item EthnicGroup: What is your ethnic group?
\item MarrigeStatus: What is your marital status now: are you married, living together with someone as if married, widowed, divorced, or separated?
\item PartnerNumber12Mo: Number of people they had sex with in the last 12 months
\item PLWHSupportGroup: Have you ever attended a support group for people living with HIV?
\item PregNum: How many times have you been pregnant including a current pregnancy?
\item RelationToHeadOfHouse: What is your relationship to the head of the household?
\item SellSexEver: Have you ever sold sex for money?
\item SyphilisTestInPreg: When you were pregnant, were you offered a test for syphilis?
\item TravelTime: At your last HIV care visit, approximately how long did it take you to travel from your home (or workplace) one way?
\item Urban: Urban Area Indicator
\item ViolenceOK?: Do you believe it is right for a man to hit or beat his wife/partner?
\item WorkLast12Mo: Have you done any work in the last 12 months for which you received a paycheck, cash or goods as payment?
\item Zone: Zone name
\end{enumerate}

\subsection{Codebook for Covariates in Figure~\ref{fig:tri90awaremale}}
\label{Codebook:tri90awaremale}


\begin{enumerate}
\item AdditionalPartner: Do you have additional spouse(s)/partner(s) that live with you?
\item AgeGroup: Age groups for population pyramid
\item AlcoholFrequency: How often do you have a drink containing alcohol?
\item BuySexEver: Have you ever paid money for sex?
\item CircumcisedStatus: Are you circumcised or planning to get circumcised?
\item EasyGetCondom: If you wanted a condom, would it be easy for you to get one?
\item PartnerAge: How old is your partner? Please give your best guess.
\item PartnerNumber12Mo: Number of people they had sex with in the last 12 months
\item PLWHSupportGroup: Have you ever attended a support group for people living with HIV?
\item Region: Region Name
\item RelationToHeadOfHouse: What is your relationship to the head of the household?
\item TravelTime: At your last HIV care visit, approximately how long did it take you to travel from your home (or workplace) one way?
\item ViolenceOK?: Do you believe it is right for a man to hit or beat his wife/partner?
\item WantMoreChild: Would you like to have a/another child?
\item WealthQuintile: Wealth quintile
\item WifeNumLiveElsewhere: How many wives/partners do you have who live elsewhere?
\item WomenCondomHaveSexALot?: Do you believe women who carry condoms have sex with a lot of men?
\item WorkLast12Mo: Have you done any work in the last 12 months for which you received a paycheck, cash or goods as payment?
\item Zone: Zone name
\end{enumerate}

\subsection{Codebook for Covariates in Figure~\ref{fig:tri90artfemale}}
\label{Codebook:tri90artfemale}


\begin{enumerate}
\item AntenatalCareLastPreg: Flag if mother who gave birth 3 years preceding survey received antenatal care during last pregnancy
\item EthnicGroup: What is your ethnic group?
\item LastChildBreastfeed: Mother's current and past breast feeding status
\item PLWHSupportGroup: Have you ever attended a support group for people living with HIV?
\item PregCurrentStatus: Are you pregnant now?
\item PregPlan: When you were pregnant, did you plan to get pregnant at that time?
\item SyphilisTestInPreg: When you were pregnant, were you offered a test for syphilis?
\item TravelTime: At your last HIV care visit, approximately how long did it take you to travel from your home (or workplace) one way?
\item Urban: Urban Area Indicator
\item ViolenceOK?: Do you believe it is right for a man to hit or beat his wife/partner?
\item WifeNumOfHusband: Including yourself, in total, how many wives or live-in partners does your husband or partner have?
\end{enumerate}

\subsection{Codebook for Covariates in Figure~\ref{fig:tri90artmale}}
\label{Codebook:tri90artmale}


\begin{enumerate}
\item AbnormPenisDischarge: During the last 12 months, have you had an abnormal discharge from your penis?
\item AnalSexEver: Have you ever had anal sex?
\item CircumcisedStatus: Are you circumcised or planning to get circumcised?
\item CondomLastPaidSex: Flag if condom was used at last paid sexual intercourse
\item FirstSexForced: The first time you had vaginal or anal sex, was it because you wanted to or because you were forced to?
\item PartnerNumber12Mo: Number of people they had sex with in the last 12 months
\item PLWHSupportGroup: Have you ever attended a support group for people living with HIV?
\item RelationToLastSexPartner: Relationship status with their last sex partner in the past 12 months
\item SeekMedicalHelp: Did you see a doctor, clinical officer or nurse because of these problems?
\item TravelTime: At your last HIV care visit, approximately how long did it take you to travel from your home (or workplace) one way?
\item WantMoreChild: Would you like to have a/another child?
\item WifeNumLiveElsewhere: How many wives/partners do you have who live elsewhere?
\item Zone: Zone name
\end{enumerate}

\subsection{Codebook for Covariates in Figure~\ref{fig:tri90vlsfemale}}
\label{Codebook:tri90vlsfemale}


\begin{enumerate}
\item AlcoholFrequency: How often do you have a drink containing alcohol?
\item CondomLastSex: Indicator for used condom at last sexual encounter in the past 12 months
\item EverWidowed: Have you ever been widowed? That is, did a spouse ever die while you were still married or living with them?
\item ForceSexTimes: How many times in your life have you been physically forced to have sex?
\item RelationshipToViolence: Relationship between you and the person who give physical violence to you.
\item SupportGroupTimes12Mo: In the last 12 months, how many times did you attend a support group?
\item TranslatorUsed: whether translator is used or not.
\item VistDoctorLast12Mo: Have you seen a doctor, clinical officer or nurse in a health facility in last 12 months?
\end{enumerate}

\subsection{Codebook for Covariates in Figure~\ref{fig:tri90vlsmale}}
\label{Codebook:tri90vlsmale}


\begin{enumerate}
\item AbnormPenisDischarge: During the last 12 months, have you had an abnormal discharge from your penis?
\item AdditionalPartner: Do you have additional spouse(s)/partner(s) that live with you?
\item PainUrinLast12Mo: During the last 12 months, have you had pain on urination?
\item SeekMedicalHelp: Did you see a doctor, clinical officer or nurse because of these problems?
\item SexTransmitDeseaseLast12Mo: In the last 12 months, did a doctor, clinical officer or nurse tell you that you had a sexually transmitted disease?
\item VerySick3MoInLast12Mo: Has name been very sick for at least 3 months during the past 12 months, that is name was too sick to work or do normal activities?
\item WifeNum: Altogether, how many wives or partners do you have?
\end{enumerate}

%
%
%

\end{document}